\pdfoutput=1
\RequirePackage{fix-cm}
\documentclass[smallcondensed]{svjour3}     
\smartqed  
\usepackage{graphicx}
\newcommand{\MZ}[1]{}
\newcommand{\mze}[1]{#1}
\usepackage{soul}
\newcommand{\mzk}[1]{}

\newcommand{\dsdk}[1]{}
\newcommand{\DSD}[1]{}
\newcommand{\FixRevTwo}[1]{#1}
\newcommand{\FixRevOne}[1]{#1}
\newcommand{\FixRevThree}[1]{#1}
\newcommand{\FixRevFour}[1]{#1}

\newcommand{\dsdkFive}[1]{}
\newcommand{\DSDFive}[1]{}
\newcommand{\FixRevFive}[1]{#1}

\usepackage{t1enc,dfadobe}
\usepackage{color}
\usepackage{graphicx}
\usepackage{caption}
\usepackage{subcaption}
\usepackage{amsmath} 
\usepackage{amssymb}
\usepackage{wasysym}
\usepackage{upquote}
\usepackage{multirow}
\usepackage{tabularx}
\usepackage{tabulary}
\usepackage{array,booktabs}
\usepackage{float}
\usepackage{wrapfig}
\usepackage{microtype}
\usepackage{multicol}
\usepackage{footmisc}
\usepackage{floatflt}
\usepackage{tikz}
\usepackage{url}
\usepackage{natbib}
\makeatletter
\def\@mb@citenamelist{cite,citep,citet,citealp,citealt,citepalias,citetalias}
\makeatother
\usepackage{multibib}
\newcites{sec}{Image References}

\usepackage{xr}
\definecolor{IntroC}{rgb}{0.5,0,0}
\definecolor{AddedNew}{rgb}{0,0, 0}
\definecolor{AddedNew2}{rgb}{0,0, 0}
\definecolor{AddedNew3}{rgb}{0,0, 0}
\definecolor{AddedNew4}{rgb}{0,0, 0}
\definecolor{Rev1}{rgb}{1.0,0.3, 0.0}
\definecolor{Rev3}{rgb}{0.0,0.6, 0.2}
\definecolor{NoteC}{rgb}{1.0,0.0, 0.0}

\DeclareGraphicsExtensions{.pdf,.png,.jpg}

\newcommand*\rfrac[2]{{}{#1}\!/{#2}}

\begin{document}

\title{Bifurcation Analysis of Reaction Diffusion Systems on Arbitrary Surfaces
}


\author{Daljit Singh J. Dhillon        \and
        Michel C. Milinkovitch  \and
        Matthias Zwicker 
}


\institute{Daljit Singh J. Dhillon \at
              Department of Computing, \\
              Imperial College London, London, United Kingdom.\\
              \email{d.dhillon@imperial.ac.uk} \\
              Note: While conducting this work, the first author was a Ph.D. student at the Institute of Computer Science, University of Bern, Bern, Switzerland.           
           \and
           Michel C. Milinkovitch \at
              Department of Genetics and Evolution, Laboratory of Artificial and Natural Evolution  (LANE),\\
              University of Geneva, Geneva, Switzerland\\
              \email{michel.milinkovitch@unige.ch}
           \and
           Matthias Zwicker \at
              Institute of Computer Science, \\
              University of Bern, Bern, Switzerland.\\
              \email{zwicker@iam.unibe.ch}           
}

\date{Received: date / Accepted: date}
\maketitle

\begin{abstract}
In this paper we present computational techniques to investigate the solutions of \FixRevThree{two-component,} nonlinear reaction-diffusion (RD) systems on arbitrary surfaces. We build on standard techniques for  linear and nonlinear analysis of RD systems, and extend them to operate on large-scale meshes for arbitrary surfaces. In particular, we use spectral techniques for a linear stability analysis to characterize \FixRevThree{and directly compose} patterns emerging from homogeneities. We develop an implementation using surface finite element methods and a numerical eigenanalysis of the Laplace-Beltrami operator on surface meshes. In addition, we describe a technique to explore solutions of the nonlinear RD equations using numerical continuation. Here, we present a multiresolution approach that allows us to trace solution branches of the nonlinear equations efficiently even for large-scale meshes. Finally, we demonstrate the working of our framework for two RD systems with applications in biological pattern formation: a Brusselator model that has been used to model pattern development on growing plant tips, and a chemotactic model for the formation of skin pigmentation patterns. While these models have been used previously on simple geometries, our framework allows us to study the impact of arbitrary geometries on emerging patterns.
\end{abstract}

\section{Introduction}
\label{sec:introduction}

Reaction-diffusion (RD) \dsdkFive{models are}\FixRevFive{is} often used to \dsdkFive{represent various mechanisms in}\FixRevFive{model the development of} biological systems, most prominently in the study of biological pattern formation. The mathematical representation of these models results in systems of nonlinear partial differential equations (PDEs)\dsdkFive{, and to understand the functioning of the underlying biological systems, it is necessary to analyse these systems of PDEs. Fundamentally, the analysis strives to answer two}\FixRevFive{. The analysis of these systems of PDEs aim at answering two fundamental} questions: \dsdkFive{First,}\FixRevFive{(i)} what are the possible solutions that satisfy the given system of PDEs, and how can these solutions be discovered systematically; and \dsdkFive{second,}\FixRevFive{(ii)} which of these solutions are \emph{stable} against minor perturbations. \dsdkFive{There are two types of}\FixRevFive{Two} strategies \dsdk{that }are commonly used to perform \dsdkFive{the}\FixRevFive{this} analysis. First, linear stability analysis \FixRevFive{can} predict\dsdkFive{s} the emergence of new patterns near trivial, homogeneous solutions (homogeneities) of the PDEs. These emergent patterns \dsdkFive{lend}\FixRevFive{correspond to sudden} qualitative changes to the state of an RD system and\FixRevFive{, hence,} constitute \emph{bifurcations} from the homogeneity. Second, nonlinear analysis \dsdkFive{attempts to find}\FixRevFive{provides} solutions of the nonlinear RD equations \FixRevFive{far }away from the homogeneous steady-states. To construct these solutions, numerical continuation techniques are used to follow continuous branches of solutions starting from initial bifurcation patterns constructed using the linear analysis. Solutions vary gradually with one of the system parameters, called the continuation parameter, along each branch.

Several existing tools allow \dsdkFive{biologists }to delineate solutions for an RD system through linear and nonlinear bifurcation analys\dsdkFive{i}\FixRevFive{e}s. However, many of them are constrained to work with only simple surface geometries such as rectangles or hemispheres, and at low resolutions. These are serious limitations \dsdkFive{considering}\FixRevFive{given that surface geometry plays an important role in pattern formation~\citep[page 108]{murray2003mathematical}, }that most of the interesting biological domains for RD systems have rather arbitrary shapes\dsdkFive{ and surface geometry plays an important role in pattern formation (Murray, 2003)}. In addition, \dsdkFive{most interesting patterns in nature}\FixRevFive{corresponding patterns} are too complex to be \dsdkFive{represented}\FixRevFive{resolved} with low-resolution meshes.

In this paper, we develop a framework to perform bifurcation analysis for generic RD systems with two components, with or without \emph{cross diffusion}, acting on arbitrary surfaces. Unlike approaches that iteratively traverse the trivial branch for a desired continuation parameter to detect bifurcation points step-by-step, 
our proposed framework uses an \emph{analysis-synthesis} approach to directly determine bifurcation points and construct emerging patterns along the trivial branch. We exploit the \emph{Hermitian} nature of the \emph{Laplace-Beltrami} (LB) operator, acting on a given arbitrary surface, which enables the computation of a spectral basis for emerging patterns. 
This allows us to derive formulae to directly compose emergent bifurcation patterns from eigenfunctions \FixRevThree{(also called eigenmodes or wavemodes)} of the LB operator. Similarly, a bifurcation point for a given bifurcation pattern can be computed directly using its eigenvalues. We discuss several boundary conditions \FixRevThree{for our framework} that are common for biological systems and ensure that the LB operator is Hermitian. Unlike \emph{detection-based} approaches, our analysis-synthesis approach avoids missing out on bifurcations due to potential failures of a \emph{test function}. In addition, our approach allows for tracing \emph{multiple} and \emph{mixed-mode} bifurcations apart from \emph{simple} bifurcations.

Our framework also supports high-resolution meshes for triangulated surface domains. We propose a \emph{progressive geometric multigrid} approach to support multiple levels of mesh resolution. We first trace a branch at the lowest resolution for the surface mesh. Next, we use a simple two-step approach to upsample a solution pattern to the next level by first improving the mesh resolution for a solution, and then resolving the solution using an improved mesh geometry. This two-step approach progressively upsamples and converges a given pattern across multiple levels. Our multiresolution approach decouples branch tracing complexity from the complexity of dealing with a \emph{large-scale} system. Unlike traditional multigrid approaches, our two-step approach does not go back and forth across multiple levels. This lends a high degree of parallelisability to our framework. 

We demonstrate the working of our framework for a \emph{Brusselator} system with \emph{zero Dirichlet} boundary conditions to study the emergent patterns of cotyledons on a conifer tip~\citep{nagata2003reaction}. We also demonstrate its working with Murray's chemotactic model for pattern formation \dsdkFive{due to}\FixRevFive{of} skin pigmentation~\citep{murray1991pigmentation},  subject to \emph{zero Neumann} boundary conditions. 
In both cases, we illustrate the influence of surface shape on pattern formation using several example geometries. Finally, we evaluate the computational performance \dsdkFive{for}\FixRevFive{of} our multiresolution branch tracing approach. In summary, our contributions include:
\begin{itemize}
\item {{A framework for analysing two-component RD systems with or without cross diffusion that supports arbitrary triangulated surface domains.}}  
\item {{A direct \emph{analysis-synthesis} approach for computing emergent patterns and locating bifurcation points along the trivial branch based on spectral analysis of the Laplace-Beltrami operator on arbitrary triangulated surface domains. 
}
}  
\item {{A progressive geometric multigrid approach supporting high-resolution FEM discretisation to determine patterns along nonlinear branches.}}  
\item{{Two case studies illustrating the effect of arbitrary geometries on pattern formation.}}  
\end{itemize}

\section{Related Work}
\label{sec:related_work}

\paragraph{Emergent Patterns near Homogeneity.}
 
Linear stability analysis with \emph{two-component} RD systems is often employed to study the emergence of \dsdkFive{a }new pattern\FixRevFive{s} from near homogeneous patternless initial conditions. 
\citet{nagata2003reaction,nagata2013reaction}
 study the relation between the shape and size of a conifer embryo and the emergent cotyledon patterns. They use a \emph{Brusselator} RD system acting on a parametric family of spherical caps while imposing zero \emph{Dirichlet} boundary conditions near homogeneity. With this, they explain how the number of emergent cotyledons is simply the selection of a spherical cap harmonic based on the radius of the conifer and its curvature~\citep{nagata2013reaction}. 
\citet{winters1990tracking} explain emergence of heterogeneous snake skin color patterns with a \emph{chemotactic} RD model with \emph{cross diffusion} \FixRevFive{(i.e., where the flux of one component is driven by gradients in the concentration of the second component)}. They perform numerical simulations on flat rectangular domains and illustrate the similarity of emergent patterns to the patterns observed in nature, for different snake species. Winters et al. use \dsdk{homogeneous}\FixRevThree{zero} \emph{Neumann} boundary conditions for their RD system.
\dsdkFive{Again}\FixRevFive{Similarly},~\citet{gambino2013pattern} \dsdkFive{present}\FixRevFive{discuss} pattern formation due to cross-diffusion for \emph{Lotka-Volterra kinetics} between two components in a 2D rectangular domain, commonly used to model predator-prey populations. 
\citet{kealy2012nonlinear} use linear stability analysis to study onset of various spatial Turing patterns for vegetation in an arid flat land using a two-component RD system. \dsdkFive{Furthermore, they}\FixRevFive{They also} apply a weak nonlinear stability analysis to predict the long-term behaviour of these emerging vegetation patterns.  
~\citet[Chapter $3$]{murray2003mathematical}
 demonstrates the effect of \FixRevFive{both }geometry and scale on \FixRevFive{the }emergen\dsdkFive{t}\FixRevFive{ce of }patterns \dsdkFive{to}\FixRevFive{under a two-component RD system and discusses the relevance of these parameters for} explain\FixRevFive{ing} \dsdkFive{several }animal coat patterns\dsdkFive{ with a two-component RD system}. He derives an analytical form for the stripe and spot patterns that emerge on a tapering cylinder representing an animal tail. Also, he presents the selection of different stripe or patchy patterns (modes) with changes in the size of a planar 2D shape representing an animal coat\dsdkFive{ cut-out}. 
Most \dsdkFive{of existing pattern emergence studies }\FixRevFive{studies on the emergence of patterns, }such as those discussed above\FixRevFive{,} are limited to simple, well-defined surface geometries with analytically defined emergent patterns. 
\FixRevFive{Recently,~\citet{tuncer2015projected} have introduced a \emph{projected Finite Elements Method} for studying pattern formation by RD systems on surfaces that can be approximated analytically and later mapped  with \emph{Lipschitz
continuity} onto a sphere. They note that studying RD systems on arbitrary surfaces is rather a ``\emph{young and emerging research area}''--- \textup{\citep{tuncer2015projected}} and that surface geometry is crucial for such studies.
Our framework extends studies of RD systems with \dsdk{and}\FixRevThree{or} without cross-diffusion to arbitrary surface domains without any geometric constraints by directly (numerically) computing emergent patterns on them.} Also, it supports several common boundary conditions \dsdkFive{for these }\FixRevFive{that arise in }biological problems such as homogeneous Dirichlet, Neumann and Robin boundary conditions. As noted \dsdk{by Murray}\FixRevThree{earlier, (arbitrary)} surface geometry plays an important role in pattern formation\dsdk{and most of the biological surfaces have rather arbitrary geometries}. Thus our framework \dsdkFive{is}\FixRevFive{serves as }an important tool for studying emergent patterns on 'real' geometries.

\paragraph{Marginal Stability Analysis.}
Marginal stability analysis is often used to study the interaction and mutual-exclusivity of two or more emergent \emph{wavemodes}. It is also used to demarcate and characterise the parameter space of an RD system. 
\citet{kealy2012nonlinear} use analytically defined marginal stability curves to demarcate regions in the parameter space with subjectively different vegetation patterns on \dsdkFive{their }arid flat lands (\FixRevFive{represented as }\FixRevThree{2D rectangles}). 
\citet{nagata2013reaction} define marginal stability curves for emergent \dsdk{patterns}\FixRevThree{wavemodes} in terms of \FixRevThree{their} corresponding eigenvalues\dsdkFive{. With several marginal stability curves, they} \FixRevFive{and} investigate the influence of the spherical cap surface geometry on \dsdkFive{emergence}\FixRevFive{the pattern} of cotyledons \FixRevFive{development}. In particular, they note that \dsdkFive{merely }changes in the curvature or size of a plant tip may cause \FixRevFive{a change in the number of }cotyledons \dsdkFive{to appear and change in their number \dsdk{even } while the}\FixRevFive{that develop despite that the concentrations of }chemical precursors are \dsdkFive{held at a }fixed\dsdkFive{ concentration}.        
Our framework generalises such marginal stability analysis to arbitrary domains. It numerically computes the eigenvalues for the \dsdk{eigenmodes}\FixRevThree{wavemodes} that constitute emergent patterns. These \dsdkFive{numerically computed }eigenvalues can then be used to plot corresponding marginal stability curves. In its full potential, our framework supports case studies with dynamically changing arbitrary shapes for marginal stability analysis as demonstrated later.         

\paragraph{Bifurcations and Branch Tracing.}
\dsdk{Bifurcation analysis and branch tracing allow a systematic partial exploration of the solution space for an RD system with nonlinear reaction terms. In particular, branch tracing helps to discover non-trivial patterns with significant amplitudes and far-away from the homogeneity.} 
Analytical solutions for branch tracing are only possible for simple \FixRevFive{surface }domains. \citet{ma2014bifurcation} express branches and patterns for a two-component \emph{Brusselator} model acting on a $1D$ straight line domain in analytical forms. They prove that\FixRevFive{,} except for the first branch along the continuation parameter dimension, all other branches are unstable.    
\citet{mendez2008population} derive analytical expressions for tracing a branch with a single component RD system to predict the survival of an isolated $1D$ patch of a population in its surrounding $1D$ hostile environment. Using stability predictions along the branch, they establish that the survival of a population at a very low or negative growth rate depends on its initial density. Instead, we support branch tracing with numerical methods in our framework.
\citet{winters1990tracking} and \citet{maini1991bifurcating} perform branch tracing numerically for their two-component RD system with cross diffusion acting on simple 2D rectangular domains. They simulate a diverse range of complex patterns with significant amplitudes for studying snakeskin pigmentations. 
\DSD{\citet{yochelis2008formation} }\dsdkFive{Yochelis et al (2008) present a study of cardiovascular calcification patterns observed to develop in laboratory cultures of vascular-derived mesenchymal stem cells. They use a two-component \emph{Gierer-Meinhardt} RD system and perform stability analysis and branch tracing numerically on a simplified \emph{periodic} $1D$ domain. Based on the insights gained from branch tracing, they label a few intervals of the bifurcation parameter with the qualitative nature of the patterns generated in those intervals (such as bounded holes, periodic states and isolated peaks). Finally they perform numeric simulations (time integration) on $2D$ domains in those parameter intervals to generate patterns that are qualitatively similar to their observations.}
 \FixRevFive{\citet{yochelis2008formation} perform numerical branch tracing for a two-component \emph{Gierer-Meinhardt} RD system on a simplified \emph{periodic} $1D$ domain to study cardiovascular calcification patterns. They use the insights gained from branch tracing to characterise the parameter space for further experiments with $2D$ domains.} 
\citet{chien2001multiple} investigate multiple modes bifurcating at a \dsdk{same}\FixRevThree{given} bifurcation point for a two-component Brusselator system subject to~\emph{Robin boundary conditions}. They demonstrate numerical continuation of multiple branches due to mode interactions for a $2D$ square domain. 
\DSD{\citet{paulau2014fundamental} }\dsdkFive{Paulau (2004) present a study of the relation between the nonlinear \emph{Schroedinger} model and a two-component RD system, a \emph{FitzHugh-Nagumo (FHN)} model, with different reaction functions for their ability to produce \emph{solitons} (self-localized solutions) in a $2D$ square domain. Specifically speaking, Paulau performed bifurcation analysis and numerical branch tracing for (i) a FHN model with quadratic-cubic reaction and (ii) a FHN model with cubic-quintic reaction, to investigate the possible solitons that these models produce as a solution and the stability of those solutions. Paulau concludes that the radially symmetric solitons upto first order which satisfy a \emph{complex} nonlinear Schroedinger equation can be transformed to the soliton solutions of a \emph{simpler} FHN model with the cubic-quintic reaction term by adjusting the FHN model parameters.}
 \FixRevFive{\citet{paulau2014fundamental} perform numerical branch tracing for a two-component \emph{FitzHugh-Nagumo (FHN)} RD system on a $2D$ planar domain to study the properties of its localised solutions, i.e. \emph{solitons}. With this, he establishes the existence and stability of certain \emph{first higher order} radially symmetric solitons with non-zero \emph{azimuthal quantum number}\footnote{i.e, characterising the zero crossings of a circumferential profile of a radially symmetric soliton.} which require the third and fifth order nonlinear reaction terms to produce them.} 
Our framework generalises such studies with two-component RD systems to arbitrary 2D surfaces.

\paragraph{Other Cases.}
In general, studies with emergent patterns, marginal stability or bifurcation analysis may deal with more than two-components~\citep{qian2001simple}, coupled layers~\citep{yang2002spatial,vasquez2013pattern}, quasi equilibrium~\citep{rozada2014stability}, advection~\citep{vasquez2013pattern,satnoianu2001parameter,madzvamuse2013moving}, a very large number~\citep{zamora2011efficient} or range~\citep{lo2012robust} of control parameters, shear-induced instability~\citep{vasquez2013pattern}, growth induced instability~\citep{madzvamuse2008stability}, nonlinear diffusion~\citep{gambino2013turing},
fractional RD~\citep{gafiychuk2009analysis} or even non-steady state (oscillatory) or \emph{travelling wave} solutions~\citep{draelants2013numerical,banerjee2012turing,wyller2007turing,qiao2006geometry,gambino2012turing}. While our framework may be used directly or with simple modifications for only \FixRevFive{a }few of these general cases, they serve as directions for future work on our framework.\DSD{Shall we move this paragraph to the last summary Section?}

\paragraph{System Discretisation, Sparsity, large-scale and Numerical Methods.}
To numerically solve RD systems for emergent patterns and bifurcation branches, we must first discretise the PDEs. Several existing techniques allow discretisation based on either \emph{discrete geometric operators}~\citep{botsch2010polygon}, \emph{finite differences} for \emph{closest-point methods}~\citep{ruuth2008simple,macdonald2011solving} or \emph{Finite Element Methods (FEM)}. \dsdkFive{Reuter et al~
compared several  \emph{discrete geometric operators} and FEMs of different orders for solving a simple PDE expressing a \emph{Laplacian eigenvalue problem} and found cubic-FEMs to outperform other approaches.} 
We employ FEMs owing to their generality and robustness and use \emph{Deal.II}~\citep{BangerthHartmannKanschat2007}, a general purpose software library that approximates arbitrary surface geometries with quadrilateral finite elements. With discretisation, a generic RD system is expressed as a linear system of equations where individual scalar elements may still be computed using nonlinear operations on discrete surface variables.  

With discretisation and linear stability analysis near homogeneity, the problem of computing emergent patterns translates into a problem of detecting bifurcations and solving a \emph{generalised eigenvalue problem} at detected bifurcation points~\citep{winters1990tracking}. \citet{seydel2010practical} discusses standard  \emph{test function} based techniques for detecting bifurcation points along any branch. These detection methods may fail, as strategies for evaluation (and construction) of test functions are empirical in nature.\dsdkFive{ While we may exploit the known nature of the bifurcations (\emph{pitchfork}, \emph{Hopf} or a \emph{turning point})}\DSDFive{\citep{seydel2010practical,cliffe2000numerical,salinger2001scalable}}\dsdkFive{to improve detection methods they may still fail occasionally.}
 In contrast to these existing detection based approaches for bifurcations on a general branch, we exploit the spectral properties of a potential pattern emergent from a trivial branch to directly compute its bifurcation point. 
Seydel also describes methods such as\dsdkFive{ \emph{tangent prediction} and} the \emph{method of parallel computation} to solve for emergent patterns along any branch for \emph{switching}~\citep[see][Chapter $5$]{seydel2010practical}. With our simplifications near homogeneity, we directly compose an emergent pattern and exploit the principles of the \emph{method of parallel computation} to switch over to a new branch.   
We  then use a \emph{pseudo arclength continuation} approach for branch tracing as presented by \citet{salinger2002loca} which is also easily scalable to large systems~\citep{salinger2001scalable}. 
   
Several natural steady state patterns are fairly complex and we need a sufficiently large system of discrete variables to study them. \dsdkFive{Often these}\FixRevFive{These} \emph{large-scale} systems result in \emph{sparse system matrices} and we need iterative algorithms such as~\emph{Krylov space methods}~\citep{gutknecht2007brief} to \dsdkFive{exploit their sparsity for solving}\FixRevFive{solve} the aforementioned eigenvalue problems\FixRevFive{~\citep{arbenz2012lecture}}\dsdkFive{ at the given large scales}.\dsdkFive{ Arbenz et al provide an elaborate account of several such algorithms.} In our framework, we use the \emph{Trilinos Project}~\citep{heroux2005overview} software libraries that implement various numerical algorithms for large-scale systems with sparse matrices on parallel processing architectures. When dealing with large-scale systems it is important to use \emph{preconditioners} to avoid numerical instabilities due to poor conditioning of matrices. 
\dsdkFive{Sala and Heroux (2005) provide a brief description of several \emph{point and block relaxation schemes} like \emph{Jacobi and Gauss-Siedel} methods and \emph{incomplete factorization schemes} like \emph{incomplete lower/upper (ILU) and incomplete Cholesky (IC)} decompositions with or without a \emph{level-of-fill} and \emph{thresholding} for preconditioning.}In our framework, we use \dsdkFive{several of these }algorithms \FixRevFive{based on \emph{incomplete factorization schemes},} as implemented in packages like \emph{IFPACK} and \emph{AztecOO} for the Trilinos libraries.

For large-scale systems under consideration, we need \emph{multigrid methods} to make iterative algorithms practicable that otherwise generate smooth but stubborn residuals with poor convergence rates~\citep[Chapter $7$]{strang2007computational}. 
Analysing computational costs for iterative Krylov subspace methods is dependent on data, convergence criteria and numerical stability, and not generalisable~\citep[page 248]{liesen2012krylov}. 
Furthermore, for non-symmetric matrices that arise with two-component RD systems\dsdkFive{ in our framework}, the convergence of popular methods like \emph{biconjugate gradient with stabilisation (Bi-CGStab)} and \emph{generalised minimal residual (GMRES)} that we use, is poorly understood~\citep{simoncini2007recent}. \dsdkFive{However, in general, multigrid methods accelerate the removal of residual errors. This is achieved by
\emph{coarsening} a grid (discretised FEM mesh in our case) to a lower resolution which  transforms smooth, \emph{low-frequency} residuals into high frequency errors.}\FixRevFive{However, multigrid methods \emph{coarsen} a grid to increase the spatial frequency of residual errors which helps to accelerate their removal~\citep[Chapter $7$]{strang2007computational}.} In practice, more than two grid levels and several passes (\emph{V-cycle}, \emph{W-cycle} or \emph{full-multigrid}~\citep[Section $7.3$]{strang2007computational}) across multiple levels are required. Also, multigrids may be classified as \emph{geometric}~\citep{landsberg2010multigrid} or \emph{algebraic}~\citep{falgout2006introduction}~\citep[Chapter $7$]{strang2007computational}, and used as preconditioners~\citep{sala2005robust} or even directly improve efficiency and robustness of branch tracing~\citep{chien2006two}. \citet{baker2011challenges} 
demonstrate that the scalability and performance of algebraic multigrids is highly dependent on the multicore platform architecture and needs expert or empirical programming efforts for optimization. In addition, tightly coupled multilevels for existing multigrid approaches make distribution and scheduling of computational workload across parallel architectures difficult, owing to interdependencies across grid levels and \emph{decomposed domain partitions}. In contrast, we propose a simple \emph{progressive geometric multigrid} approach that decouples branch tracing from scale improvements to allow a highly parallelisable implementation. Our approach is based on a multigrid continuation approach by~\citet{bank1986pltmgc} where branch tracing is performed at the coarsest level and resultant solutions are refined for higher resolution meshes. We adapt this approach to perform alternative resolution and geometric improvements iteratively and demonstrate its working for a very high mesh resolution ($1M$ FEM nodes) for an arbitrary surface. The inherent benefits of this decoupled progressive approach includes selective and parallel refinements of branch solutions.

\section{Direct Linear Analysis with Laplacian Eigenbasis}
\label{sec:formulation}
In this section, we formulate our direct approach to compute bifurcation patterns using eigenfunctions of the Laplacian-Beltrami operator. We derive a general form for an emergent pattern near homogeneity for a generic two-component RD system acting on arbitrary surfaces, with or without cross-diffusion. Also, we derive the constraints to be satisfied by the continuation parameter for locating a bifurcation point. 
We first perform a simplification of the generic RD system equations near homogeneity into a linear form, to be satisfied by an emergent pattern (Section~\ref{ssec:linearise_rd}). Next, we present a spectral decomposition of potential patterns and the boundary conditions that allow expressing these patterns with orthogonal basis functions (Section~\ref{ssec:spectral_bc}). Then, we substitute a spectrally decomposed potential pattern into the linearised system equations to obtain an explicit general form for an emergent pattern along with expressions and conditions for its spectral coefficients (Section~\ref{ssec:bif_pattern}). We also define the bifurcation point in terms of spectral eigenvalues and system parameters. 
We discuss three cases of \emph{simple}, \emph{multiple} and \emph{mixed-mode bifurcations} and the constraints that they impose on our general derivations, and we discuss how to directly compose bifurcation patterns for these cases (Section~\ref{ssec:locating_bif}). Also, we discuss common applications such as \dsdk{(exclusive)}\FixRevThree{\emph{exclusive}} mode selection with isotropic domain-growth \dsdk{and pattern emergence under competition}.         


\subsection{Linearising Generic Two-Component RD Systems} 
\label{ssec:linearise_rd}
Let us consider a two-component general RD system with cross diffusion, defined over an arbitrary surface. It can be expressed mathematically as
\begin{align}
\frac{\partial a}{\partial t} &= \nabla^2\left[ \left({}^aD_a + {}^aD_\alpha a + {}^aD_\beta b \right) a \right] + f(a,b) \,,\nonumber\\ 
\frac{\partial b}{\partial t} &= \nabla^2\left[ \left({}^bD_b + {}^bD_\alpha a + {}^bD_\beta b \right) b \right] + g(a,b) \,. 
\label{eqn:gen_cross_rd}
\end{align}
\FixRevOne{Here, $a:\Omega \mapsto \mathbb{R}$ and $b:\Omega \mapsto \mathbb{R}$ are the concentrations of two components over the surface $\Omega$}, diffusion is represented with the Laplace-Beltrami operator $\nabla^2$, and functions $f$ and $g$ represent nonlinear reaction terms. The scalar coefficients ${}^aD_a$ and ${}^bD_b$ are positive diffusion rates, ${}^aD_\alpha$ and ${}^bD_\beta$ are non-negative \emph{self-diffusion} factors and ${}^aD_\beta$ and ${}^bD_\alpha$ are non-negative \emph{cross-diffusion} factors for the system~\citep{lou1996diffusion}.
Throughout this paper, we consider only steady state solutions of RD systems. For Equation~\ref{eqn:gen_cross_rd}, this implies that we are interested in solutions with  $\rfrac{\partial a}{\partial t}  = \rfrac{\partial b}{\partial t} = 0$.
To linearise the RD system defined in Equation~\ref{eqn:gen_cross_rd}, we first define its homogeneous steady state $(a_0,b_0)$ as a solution to the simultaneous equations $f(a_0,b_0) = 0$ and $g(a_0,b_0) = 0$. Note that depending on the complexity of $f$ and $g$ (say polynomial order), there may be multiple choices for  the homogeneous steady state $(a_0,b_0)$. Given a steady state $(a_0,b_0)$, we now perform a Taylor series expansion for the nonlinear reaction terms $f$ and $g$ for infinitesimal deviations $u = \left.\Delta a\right|_{a_0}$ and $v = \left.\Delta b\right|_{b_0}$,
\begin{align}
f(a_0 + u, b_0 + v) &=  f(a_0,b_0) +  u \left.\frac{\partial f}{\partial a}\right|_{(a_0,b_0)} + v \left.\frac{\partial f}{\partial b}\right|_{(a_0, b_0)} + n_f(u,v), \nonumber \\ 
g(a_0 + u, b_0 + v) &= g(a_0,b_0) +  u \left.\frac{\partial g}{\partial a}\right|_{(a_0,b_0)} + v \left.\frac{\partial g}{\partial b}\right|_{(a_0, b_0)} + n_g(u,v),
\label{eqn:taylor_series}
\end{align}
where $n_f$ and $n_g$ are polynomial functions containing the second and higher order terms in $u$ and $v$ for their respective Taylor series expansions. In other words,  $n_f$ and $n_g$ represent the nonlinear part of the reaction terms for the system defined by Equation~\ref{eqn:gen_cross_rd} near \dsdk{steady state pattern}\FixRevThree{the homogeneous steady state} $(a_0, b_0)$.
Now substituting $a=a_0 + u$, $b=a_0 + v$, $f(a_0 + u, b_0 + v)$, and $g(a_0 + u, b_0 + v)$ from Equation~\ref{eqn:taylor_series} into Equation~\ref{eqn:gen_cross_rd} and ignoring the nonlinear terms yields 

\noindent\begin{minipage}{0.54\textwidth}
\begin{align}
\frac{\partial u}{\partial t} &= {}^uD_u \nabla^2 u \,\,+\,{}^uD_v \nabla^2 v \,\,+\,\, f_l(u,v) \, , \nonumber\\ 
\frac{\partial v}{\partial t} &= {}^vD_v \nabla^2 v \,\,+\,{}^vD_u \nabla^2 u \,\,+\,\, g_l(u,v) \, ,\quad   
\label{eqn:gen_cross_rd_uv_lin}
\end{align}\vspace{0.1cm}
\end{minipage}
\hfill{}$\Biggr|$
\begin{minipage}{0.42\textwidth}
\begin{align} \break
\text{with }\quad\quad&\nonumber\\ f_l(u,v) &= \,\, {}^uK_u\, u \,\,+\,\, {}^uK_v\, v  \, , \nonumber\\
 g_l(u,v) &= \,\, {}^vK_u\, u \,\,+\,\, {}^vK_v\, v  \, ,  
\end{align}\vspace{0.1cm}
\end{minipage} 

\FixRevOne{\noindent for $u:\Omega\mapsto\mathbb{R}$ and $v:\Omega\mapsto\mathbb{R}$.
Here, new diffusion coefficients $\lbrace {}^uD_u,{}^uD_v, {}^vD_u, {}^vD_v\rbrace$ and reaction coefficients $\lbrace {}^uK_u,{}^uK_v, {}^vK_u, {}^vK_v\rbrace$ are defined in terms of old coefficients in Equation~\ref{eqn:gen_cross_rd}, $a_0, b_0$, and partial derivatives $\rfrac{\partial f}{\partial a}$, $ \rfrac{\partial f}{\partial b}$, $ \rfrac{\partial g}{\partial a}$ and $\rfrac{\partial g}{\partial b}$ evaluated at $(a_0,b_0)$. }See the supplemental material (SM01.D1) for the definition of these new coefficients along with necessary derivations. We emphasize that for the above linearisation, all non-linear terms in $u$, $v$, $\nabla u$, $\nabla v$, $\nabla^2 u$ and $\nabla^2 v$ are considered to be negligible and ignored since $|u| \ll a_0$ and $|v| \ll b_0$ near homogeneity. 
  
\subsection{Spectral Decomposition and Boundary Conditions}
\label{ssec:spectral_bc}
\FixRevOne{
Our framework performs bifurcation analysis near homogeneity using spectral analysis of the Laplace-Beltrami operator $\nabla^2$. Any surface function such as $u$ and $v$ in Equation~\ref{eqn:gen_cross_rd_uv_lin} can be expressed in terms of eigenmodes of $\nabla^2$. Also, if a second-order linear operator such as $\nabla^2$ is \emph{Hermitian}, then its eigenmodes form a set of orthonormal basis functions. 
Using an orthonormal spectral decomposition, we derive the form and conditions for an emergent pattern directly in terms of the eigenmodes and eigenvalues for the Laplacian-Beltrami operator. 

Now, in order to ensure the orthonormality of the basis functions, we need to consider the boundary conditions that $u$ and $v$ are subject to. Most biological problems expressed as RD systems are subject to either periodic \FixRevThree{boundary conditions} or \dsdk{homogeneous}\FixRevThree{zero} \emph{Dirichlet}, \emph{Neumann}, or \emph{Robin} boundary conditions near homogeneity, or they deal with \emph{closed} surfaces without boundaries. We show in the supplemental material (SM01.D2) that all these cases satisfy the Hermitian property of the Laplace-Beltrami operator. 

\paragraph{Spectral decomposition.}
To generate an orthonormal basis set $\lbrace\phi_k\rbrace$ using the Laplace-Beltrami operator $\nabla^2$, we must solve the corresponding eigenvalue problem 
\begin{align}
\nabla^2 \phi_k = - \lambda_k \phi_k\,,\quad\forall k\,.
\label{eqn:laplacian_basis}
\end{align}
Here, all the eigenvalues $\lambda_k$ are real and non-negative since $\nabla^2$ is Hermitian. We thus assume that the basis functions form an ordered set, where the ordering index $k$ satisfies $\lambda_k \leq \lambda_{k+1}$. Using the basis functions $\lbrace\phi_k\rbrace$ we can express a smooth surface function $f$ as
\begin{align}
f = \sum_{k} {f_k\,\phi_k}\, , \text{\hspace{0.5cm}where\hspace{0.5cm}} f_k = \langle f, \phi_k \rangle\,\text{\hspace{0.5cm}} \forall k\,,
\label{eqn:spectral_decompose}
\end{align}
and $f_k$ are called spectral coefficients. Next, we leverage the spectral decomposition in Equation~\ref{eqn:spectral_decompose} to express a pattern emergent near homogeneity for a generic RD system defined in Equation~\ref{eqn:gen_cross_rd}.
}

\subsection{Bifurcation Patterns near Homogeneity}
\label{ssec:bif_pattern}
\FixRevOne{While the spectral decomposition suggests that potential patterns could in general contain any superposition of eigenmodes, the conditions near homogeneity impose additional constraints on actual emergent patterns. } We now derive an as--general--as--possible form for emergent steady-state patterns of our generic RD system formulation that respects these constraints. 
 
Consider a steady-state bifurcation pattern $(u^b, v^b)$ \FixRevOne{, where the superscript $b$ denotes that this emergent pattern is a bifurcation from the trivial homogeneous solution}. Substituting its spectral decomposition from Equation~\ref{eqn:spectral_decompose} into Equation~\ref{eqn:gen_cross_rd_uv_lin}, and setting the temporal derivatives to zero to obtain a steady-state solution, gives us
\begin{align}
\quad\frac{\partial u^b}{\partial t} &= {}^uD_u \nabla^2 \sum_k{u_k \phi_k} \,\,+\,\, {}^uD_v \nabla^2 \sum_k{v_k \phi_k} \,\,+\,\, {}^uK_u\, \sum_k{u_k \phi_k} \,\,+\,\, {}^uK_v\, \sum_k{v_k \phi_k}  \,\, &= \,\,0\, ,&\text{\hspace{05cm} }\nonumber\\
\quad\frac{\partial v^b}{\partial t} &= {}^vD_v \nabla^2 \sum_k{v_k \phi_k} \,\,+\,\,{}^vD_u \nabla^2 \sum_k{u_k \phi_k} \,\,+\,\, {}^vK_u\, \sum_k{u_k \phi_k} \,\,+\,\, {}^vK_v\, \sum_k{v_k \phi_k}  \,\, &= \,\,0\, .&
\label{eqn:bif_solution}
\end{align}  
For simplicity, we dropped the superscript $b$ from the spectral coefficients $u_k$ and $v_k$.
Simplifying Equation~\ref{eqn:bif_solution} using substitutions $\nabla^2\phi_k=-\lambda_k \phi_k$, $\forall k$, and imposing linear independence of orthonormal basis functions $\phi_k$ yields the following relations between the spectral coefficients $u_k$, $v_k$, and eigenvalues $\lambda_k$, $\forall k$ (see supplemental material (SM01.D3) for a detailed derivation),
\begin{align}
 ({}^uK_u\,-\, {}^uD_u \lambda_k)\, u_k  \,\,+\,\, ({}^uK_v\, - {}^uD_v \lambda_k) \, v_k \,\, &= \,\,0\, , \nonumber\\
 ({}^vK_v\,-\, {}^vD_v \lambda_k)\, v_k  \,\,+\,\, ({}^vK_u\, - {}^vD_u \lambda_k) \, u_k \,\, &= \,\,0\,.
\label{eqn:bif_constraints_cross_diff}
\end{align}
For each non-zero pair of $u_k$ and $v_k$, we derive the following constraint on the corresponding eigenvalue $\lambda_k$ from the above equation,
\begin{align}
 ({}^uD_u {}^vD_v \,\,-\,\, {}^vD_u {}^vD_u)&  \lambda_k^2 \,\,-\,\, \text{\hfill{}} &\nonumber\\ 
 ({}^uD_u {}^vK_v \,\,+\,\, {}^vD_v &{}^uK_u \,\,-\,\,{}^uD_v {}^vK_u\,\,-\,\,{}^vD_u {}^uK_v ) \lambda_k \,\,+ \,\, 
  {}^uK_u {}^vK_v - {}^uK_v {}^v K_u \,\, = \,\,0 \,. \label{eqn:lambda_constraints}
\end{align} 
{\color{AddedNew}
Equation~\ref{eqn:lambda_constraints} is quadratic in $\lambda_k$ and it admits at most two real valued roots for $\lambda_k$, say $\Lambda_m$ and $\Lambda_n$. 
This implies that at most two set of \FixRevOne{$\lbrace \phi_i \rbrace = \lbrace \phi_k \,|\, \lambda_k = \Lambda_m \rbrace$ and $\lbrace \phi_j \rbrace = \lbrace \phi_k \,|\, \lambda_k = \Lambda_n \rbrace$} may be combined linearly to form an emergent pattern. 
Thus the most general form for a bifurcation pattern emergent near homogeneity for an RD system as in Equation~\ref{eqn:gen_cross_rd_uv_lin} is
\begin{align} 
\label{eqn:gen_steady_state_sol}
u^b = \sum_{\lbrace i \rbrace} {u_i \phi_i} \,\, + \,\,\sum_{\lbrace j \rbrace} {u_j \phi_j}\,,\quad\quad
v^b = \sum_{\lbrace i \rbrace} {v_i \phi_i}  \,\,+\,\, \sum_{\lbrace j \rbrace} {v_j \phi_j}\,,\,
\nonumber\\
\text{with}\quad \lbrace i \rbrace\,=\, \lbrace k \,|\,\lambda_k=\Lambda_m\rbrace\, \quad\text{and}\quad \lbrace j \rbrace\,=\, \lbrace k \,|\,\lambda_k=\Lambda_n\rbrace\,. 
\end{align}
Now, an important requirement for patterns given by Equation~\ref{eqn:gen_steady_state_sol} to emerge due to diffusion driven instabilities is that the RD systems must be linearly stable in absence of diffusion. Murray expresses this requirement in terms of the differentials of the reaction terms in an RD system equation~\citet[see][Equation $2.19$]{murray2003mathematical}. For RD systems given by Equation~\ref{eqn:gen_cross_rd}, near homogeneity, the stability constraints are expressed in terms of system parameters as
\begin{align}
\frac{\partial f_l}{\partial u} \,+\, \frac{\partial g_l}{\partial v} \,= \, {}^uK_u \,+\,{}^vK_v \, < 0\,, \text{ and}\quad
\frac{\partial f_l}{\partial u} \, \frac{\partial g_l}{\partial v} \,-\,
\frac{\partial f_l}{\partial v} \, \frac{\partial g_l}{\partial u}\,= \, {}^uK_u \,{}^vK_v \,-\, {}^uK_v \,{}^vK_u \, > 0\,.
\label{eqn:linear_stability_preconditions}
\end{align}

{
\begin{figure}[t]
\centering
\begin{tikzpicture}
\node[anchor=south west,inner sep=0] (image) at (0,0) {
\includegraphics[width=0.9\textwidth]{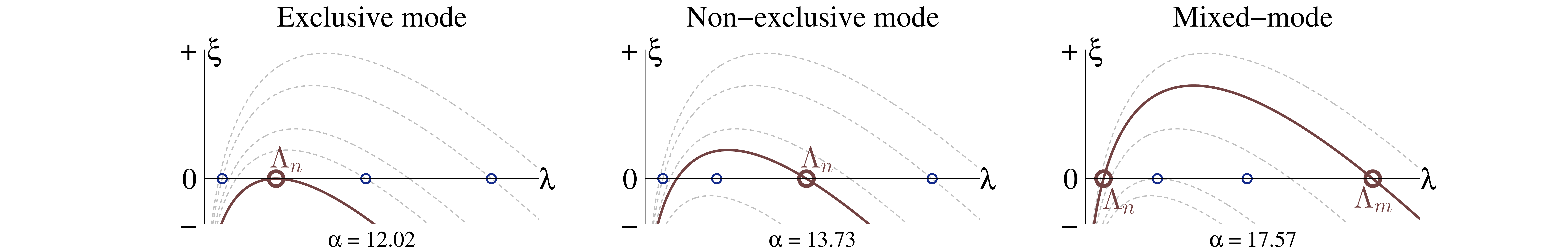}
};
\begin{scope}[x={(image.south east)},y={(image.north west)}]
    \end{scope}
\end{tikzpicture}  
\caption{Dispersion relation curves for the first three bifurcations for Murray's {chemotactic} RD system~\citep{winters1990tracking} with $\alpha$ as a bifurcation parameter. Potential eigenvalues are represented along the $\lambda$-axis and the corresponding temporal growth rates are represented along the $\xi$-axis. Actual eigenvalues of the LB operator are shown as circles. As we move along the trivial branch by increasing the value of the bifurcation parameter $\alpha$, different eigenmodes (represented by bigger brown circles) become unstable to branch out new bifurcations.}
\label{fig:dispersion_relations}
\end{figure}
To facilitate direct composition of emergent patterns we classify them based on two criteria. First, we classify patterns as (i) \emph{exclusive mode selections} or (ii) \emph{non-exclusive mode selections} based on certain conditions for the diffusion induced instability. Diffusion driven instabilities may activate multiple eigenmodes $\phi_k$ to grow or emerge simultaneously under a fixed set of system parameters. This is captured by the dispersion relation~\citep[page $86$]{murray2003mathematical}\MZ{add citation here}, which indicates the range of eigenvalues that are unstable for the fixed system parameters. The dispersion relation provides a \emph{growth rate} $\xi_k$ for each eigenvalue $\lambda_k$, and only eigenmodes with non-negative growth rates are unstable and may participate in pattern formation. 

\dsdk{The figure above}\FixRevThree{Figure~\ref{fig:dispersion_relations}}\MZ{make one figure with all three plots} plots dispersion relations as curves of growth rates $\xi$ over potential eigenvalues $\lambda$. The shapes of the curves are determined by the given system parameters, and varying a free continuation parameter leads to a family of curves as shown here. Circles along the $\lambda$--axis indicate actual eigenvalues $\lambda_k$ from the spectral analysis. In exclusive mode selection, eigenmodes corresponding to exactly one eigenvalue, say $\Lambda_n$, become unstable (they have a non-negative growth rate) at the bifurcation point. In non-exclusive mode selection, the system becomes unstable to a new set of eigenmodes with eigenvalue $\Lambda_n$, while it remains unstable to other modes with positive growth rates. 

Further, similar to \citet{chien2001multiple}\footnote{\FixRevThree{While \citet{chien2001multiple} label all bifurcations made of two or more eigenmodes as \emph{mixed-mode}, we reserve this term only for the bifurcations that involve eigenmodes with different eigenvalues.}}, we classify bifurcations as: (a) \emph{simple} if only a single wavemode constitutes the emergent pattern, (b) \emph{multiple} if more than one wavemode constitutes the emergent pattern but all such wavemodes have the same eigenvalue, and (c) \emph{mixed-mode} for the bifurcations with constituent wavemodes of more than one eigenfrequency, say $\Lambda_n$ and $\Lambda_m$. \dsdk{The figure on the right}\FixRevThree{Figure~\ref{fig:dispersion_relations}} on the right illustrates the dispersion relation for a mixed-mode pattern at its bifurcation point.  
Next, we derive different formulae to compose bifurcation patterns directly for these cases.

}
\subsection{Composing Bifurcation Patterns for Continuation}
\label{ssec:locating_bif}

Building on the derivations from Sections~\ref{ssec:linearise_rd},~\ref{ssec:spectral_bc} and~\ref{ssec:bif_pattern}, we now derive equations to specifically compose simple, multiple, and mixed-mode bifurcations for the RD system in Equation~\ref{eqn:gen_cross_rd} (near homogeneity), both under exclusive and non-exclusive mode selection.

\subsubsection{Simple Bifurcations}
\label{sssec:simple_bif}
Simple bifurcation patterns are composed of a single wave $\phi_i$ such that the algebraic multiplicity of its corresponding eigenvalue $\lambda_i = \Lambda$ is $1$. Let for some $i$, $(u^b = u_i \phi_i\,,\,v^b = v_i \phi_i)$ be the steady state bifurcation pattern of interest. We now investigate the conditions that the system parameters must satisfy to branch out a steady state pattern $(u^b,\,v^b)$. 

{\color{AddedNew}
\paragraph{Exclusive Mode Selection.}
For simple bifurcations in exclusive mode selection problems, the \emph{quadratic} Equation~\ref{eqn:lambda_constraints} must have only one real root for $\lambda_i = \Lambda$ satisfying
\begin{align}
2\,\Lambda\,({}^uD_u {}^vD_v \,\,-\,\, {}^vD_u {}^vD_u) \,\,=\,\,{}^uD_u {}^vK_v \,\,+\,\, {}^vD_v {}^uK_u \,\,-\,\,{}^uD_v {}^vK_u\,\,-\,\,{}^vD_u {}^uK_v
\label{eqn:cross_exclusive_lambda}
\end{align}
at the bifurcation point. For RD systems without cross-diffusion, ${}^uD_v = {}^vD_u =0$ and the above relation becomes $2\,\Lambda\,{}^uD_u {}^vD_v \,\,=\,\,{}^uD_u {}^vK_v \,\,+\,\, {}^vD_v {}^uK_u\,$. Thus for studying exclusive mode selection for RD systems without cross-diffusion, the continuation parameter must be associated with either ${}^uK_u$ or ${}^vK_v$.

An interesting class of problems deals with isotropic growth of the domain as the mode selection criterion~\citep[page $151$]{murray2003mathematical}. For such cases, let us introduce  a common scale factor $\gamma > 0$ for all the parameters to represent isotropic growth~\footnote{\dsdk{The logic behind this common scale factor is that reaction rates are inversely proportional to surface area which increases with growth and the balance between reaction and diffusion terms is effected. Refer Murray's book}\FixRevThree{See Murray's book~\citep[page $78$]{murray2003mathematical} for an interpretation of the scale factor $\gamma$.}\dsdk{for details.}}, i.e. ${}^uK_u =\gamma\,\,\,{}^uG_u$, ${}^uK_v =\gamma\,\,\,{}^uG_v$, ${}^vK_u =\gamma\,\,\,{}^vG_u$ and ${}^vK_v =\gamma\,\,\,{}^vG_v$. Substituting these terms in Equation~\ref{eqn:cross_exclusive_lambda} gives us $\gamma$ at the bifurcation point directly in terms of the system parameters as
\begin{align}
\gamma\,\,=\,\,\rfrac{2\,\Lambda\,({}^uD_u {}^vD_v \,\,-\,\, {}^vD_u {}^vD_u)\,\,}
{\,\left({}^uD_u {}^vG_v \,\,+\,\, {}^vD_v {}^uG_u \,\,-\,\,{}^uD_v {}^vG_u\,\,-\,\,{}^vD_u {}^uG_v \right)}\,. 
\label{eqn:scale_factor_cross}
\end{align}

\paragraph{Non-Exclusive Mode Selection.}
For simple bifurcations that emerge non-exclusively, the $\lambda_i = \Lambda$ for a given wavemode $\phi_i$ may not be a unique root for Equation~\ref{eqn:lambda_constraints}. Nevertheless, for detecting bifurcations near homogeneity with all but one unknown system parameter, we can use Equation~\ref{eqn:lambda_constraints} with $\lambda_i = \Lambda$ to directly compute the free (continuation) parameter. 

Thus, for all simple bifurcations along the trivial branch 
, we can directly compute a continuation parameter to locate the bifurcation point for a given wavemode $\phi_i$ by substituting $\Lambda=\lambda_i$ in either Equation~\ref{eqn:cross_exclusive_lambda} (exclusive mode selection),~\ref{eqn:scale_factor_cross} (exclusive mode selection under domain growth) or~\ref{eqn:lambda_constraints} (non-exclusive mode selection). Finally, with all system parameters known, we can obtain the spectral coefficients $u_i$ and $v_i$ by solving the simultaneous Equations~\ref{eqn:bif_constraints_cross_diff}. This yields the desired bifurcation pattern $(u^b,v^b)$, and we can switch to the new branch for tracing (see Section~\ref{sssec:BranchTracing})\footnote{Note that Equation~\ref{eqn:bif_constraints_cross_diff} is homogeneous and the terms $u_i$ and $v_i$ can only be solved upto a common scale factor, say $s_i = v_i/u_i$.}. 
}
     
\subsubsection{Multiple Bifurcations}
\label{sssec:multiple_bif}
{
For an eigenvalue $\lambda_i = \Lambda$ with algebraic multiplicity $> 1$, we have multiple candidate wavemodes $\phi_i$ that satisfy the conditions and equations that we derived for simple bifurcations. Thus, for such cases, every linear combination of these wavemodes is a bifurcation pattern emergent at the same bifurcation point located using the results from \dsdk{previous sub-section}\FixRevThree{Section~\ref{sssec:simple_bif}}. We discuss branch switching and continuation for such multiple bifurcations in more detail in Section~\ref{sec:framework}.}

\subsubsection{Mixed-mode Bifurcations}
\label{sssec:mixedmode_bif}
{\color{AddedNew}
Let us now consider an \emph{as--general--as--possible} emergent steady state bifurcation pattern as given in Equation~\ref{eqn:gen_steady_state_sol}. 
This means there are eigenmodes of two different eigenvalues $\Lambda_m$ and $\Lambda_n$, and we assume that $\Lambda_{m} < \Lambda_{n}$. 
Let us define $s_i = v_i/u_i$ and  $s_j = v_j/u_j$. Now, since $\lambda_i = \Lambda_m$, $\forall i$, Equation~\ref{eqn:bif_constraints_cross_diff} implies that all $s_i$ are equal, say $s_i = s_m$, $\forall i$. Similarly, $s_j = s_n$ (say), $\forall j$. 
Substituting these scale factors with their respective eigenvalues in Equation~\ref{eqn:bif_constraints_cross_diff} yields\footnote{See our supplemental material (SM01.D4) for detailed derivation for these coefficients and Equation~\ref{eqn:cross_diffusion_condition}.}
\begin{align}
{}^u K_u &= \rfrac{\left[\,{}^uD_u  (s_n \Lambda_{m} - s_m \Lambda_{n})\,\,+\,\,
{}^uD_v s_m s_n  (\Lambda_{m} - \Lambda_{n})\,\right]\,\,}{\,(s_n - s_m)} \,,\nonumber\\
{}^u K_v &= \rfrac{\left[\, {}^uD_u (\Lambda_{n} - \Lambda_{m})\,\,+\,\,
{}^uD_v (s_n \Lambda_{n} - s_m \Lambda_{m})\,\right]\,\,}{\,(s_n - s_m)} \,,\nonumber\\
{}^v K_u &= \rfrac{\left[\,{}^vD_v s_m s_n (\Lambda_{m} - \Lambda_{n})\,\,+\,\,
{}^vD_u (s_n\Lambda_{m} - s_m\Lambda_{n})
\,\right]\,\,}{\,(s_n - s_m)} \,,\nonumber\\
{}^v K_v &= \rfrac{\left[\, {}^vD_v (s_n \Lambda_{n} - s_m \Lambda_{m})\,\,+\,\,
{}^vD_u (\Lambda_{n} - \Lambda_{m})\,\right]\,\,}{\,(s_n - s_m)} \,.
\label{eqn:mixedmode_params}
\end{align}  
Now, one of the linear stability requirements in absence of diffusion is that\dsdkFive{${}^u K_u + {}^v K_v < 0$} \FixRevFive{${}^uK_u \,{}^vK_v \,> \, {}^uK_v \,{}^vK_u $} (Equation~\ref{eqn:linear_stability_preconditions}). Expanding this inequality with substitutions from Equation~\ref{eqn:mixedmode_params} gives us a constraint on the diffusion parameters,
\begin{align}
{}^uD_u {}^vD_v > {}^uD_v {}^vD_u\,,
\label{eqn:cross_diffusion_condition}
\end{align}
which needs to be satisfied to obtain a mixed mode bifurcation. 
Next, using \FixRevThree{the fact that} $\Lambda_n$ and $\Lambda_m$ \dsdk{in}\FixRevThree{are solutions for $\lambda_k$ in} Equation~\ref{eqn:lambda_constraints}, we \dsdk{obtain the relation}\FixRevThree{can solve for the continuation parameter which is the only unknown in the following equation,}
\begin{align}
({}^uD_u {}^vD_v \,\,-\,\, {}^vD_u {}^vD_u)^2(\Lambda_n - \Lambda_m)^2 \,+&\,4\,({}^uD_u {}^vD_v \,\,-\,\, {}^vD_u {}^vD_u)\,({}^uK_u {}^vK_v - {}^uK_v {}^v K_u) \,-\nonumber\\
({}^uD_u {}^vK_v \,\,+\,\, {}^vD_v &{}^uK_u \,\,-\,\,{}^uD_v {}^vK_u\,\,-\,\,{}^vD_u {}^uK_v )^2 = 0\,.& 
\label{eqn:mixedmode_bifpoint}
\end{align} 
\DSD{Following two sentences are repeated later and thus redundant here. \st{With all but one system parameter known, Equation}~\ref{eqn:mixedmode_bifpoint} \st{becomes a quadratic for the unknown continuation parameter that we can solve for easily. The solution for Equation}~\ref{eqn:mixedmode_bifpoint} \st{must be a real number and satisfy the linear stability constraints in Equation}~\ref{eqn:linear_stability_preconditions}.} Finally, with all system parameters determined, we can compute $s_m$ and $s_n$ as
\begin{align}
s_m\,&=\,\rfrac{({}^uD_u\Lambda_m -{}^uK_u)}{({}^uK_v - {}^uD_v\Lambda_m)}\,,\nonumber\\
s_n\,&=\,\rfrac{({}^vD_u\Lambda_n -{}^vK_u)}{({}^vK_v - {}^vD_v\Lambda_n)}\,.
\label{eqn:mixedmode_sm_sn}
\end{align}

\FixRevOne{In summary, to study a mixed-mode bifurcation we start composing a desired emergent pattern by selecting two eigenvalues, $\Lambda_m$ and $\Lambda_n$,  and the corresponding sets of eigenmodes $\lbrace\phi_i \,|\, \lambda_i = \Lambda_m\rbrace$ and $\lbrace\phi_j \,|\, \lambda_j = \Lambda_n\rbrace$. In addition, we freely choose desired spectral coefficients $u_i$ and $u_j$. 
We also set values for all but one system parameter such that they satisfy Equation~\ref{eqn:cross_diffusion_condition}. Then, we compute the unknown continuation parameter by solving the (quadratic) Equation~\ref{eqn:mixedmode_bifpoint}. \FixRevOne{In order to continue with branch tracing, the solved bifurcation parameter must be real valued and satisfy preconditions in Equation~\ref{eqn:linear_stability_preconditions}. Else, our framework reports an error.} 
Finally, we compute $s_m$ and $s_n$ using Equation~\ref{eqn:mixedmode_sm_sn}, and the spectral coefficients $v_i$ and $v_j$ using  $u_i$, $u_j$, $s_m$ and $s_n$. Now, all terms are determined to compose the bifurcation pattern $(u^b,v^b)$ using Equation~\ref{eqn:gen_steady_state_sol}.
}

\section{Numerical Method}
\label{sec:framework}
{\color{AddedNew3}
In this section we describe the numerical implementation of our framework in more detail. We build on the \emph{Trilinos} and \emph{Deal.II} libraries to provide an implementation of our \emph{proposed method} for bifurcation analysis near homogeneity. 
Our framework uses FEM-based surface discretisation using quadrilateral finite elements (Q-FEs), and the order of the FEs can be configured up to three. 
We support the following operations for bifurcation analysis and branch tracing: approximating Laplacian eigenfunctions (Section~\ref{sssec:ComputingEigFn}),  resolving bifurcations (Section~\ref{sssec:ComposeBifPat}),  and branch tracing (Section~\ref{sssec:BranchTracing}). In addition, we will describe a strategy for resolving patterns at higher resolutions in Section~\ref{sec:adaptation}. 
Our framework also includes a generic, indirect \emph{reference method} for branch detection using a \emph{test function} \citep[see][Section $5.3$]{seydel2010practical}. We use it for various comparisons during the evaluation of our framework.

\subsection{Approximating Laplacian Eigenfunctions}
\label{sssec:ComputingEigFn}

We saw in Section~\ref{ssec:bif_pattern} that the eigenfunctions $\phi_k$ of the Laplacian of a surface domain $\Omega$ are the building blocks for composing bifurcation patterns, given an RD system satisfying one of the boundary conditions discussed in Section~\ref{ssec:spectral_bc}. Hence computing these eigenfunctions is a core functionality of our approach. Interestingly, the eigenfunctions depend only on the shape of the surface domain $\Omega$ and the boundary conditions\dsdk{, but}\FixRevThree{. They} are independent, however, of the RD system formulation and its parameters. We use an FEM discretisation of the domain $\Omega$ and apply a \emph{Galerkin method} to discretise the eigenvalue problem in Equation~\ref{eqn:laplacian_basis} as
\begin{align}
\mathbf{L} \mathbf{b}_k =\,\,-\lambda_k \mathbf{M} \mathbf{b}_k\,.
\label{eqn:DisreteLapGenEigProblem}
\end{align}
Here, $\mathbf{b}_k$ is a discrete vector representation of the eigenfunction $\phi_k$. We obtain the \emph{stiffness matrix} $\mathbf{L}$ and the \emph{mass matrix} $\mathbf{M}$ by applying a \emph{weak formulation} integration to the Laplacian operator $\nabla^2$ \FixRevThree{and the eigenfunction $\phi_k$, respectively}. As before, $\lambda_k$ is the eigenvalue corresponding to the eigenfunction $\phi_k$. 
Equation~\ref{eqn:DisreteLapGenEigProblem} is a generalised eigenvalue problem, and in our case we obtain a large-scale system of sparse matrices. 
We use the \emph{Anasazi} eigensolver package from the Trilinos library for finding the eigenvectors $\mathbf{b}_k$. For large-scale general eigenvalue problems a \emph{shift-invert} approach is commonly used to solve for a \emph{band} of eigenvectors as recommended by~\citet{levy2010spectral}. However, for RD systems with zero Neumann boundary conditions, $\mathbf{L}$ is singular and the shift-invert method cannot be used to compute the lowest frequency band of eigenvectors. To keep things simple, we apply $\mathbf{M}^{-1}$ as a preconditioner to both sides and solve the resulting standard eigenvalue problem. We avoid explicit computation of the possibly non-sparse, large matrix $\mathbf{M}^{-1}$ with the use of the \emph{AztecOO} package from the Trilinos library. AztecOO provides an inner loop implementation for each application of matrix $\mathbf{M}^{-1}$ to a vector (say) $\mathbf{y}$ for computing $\mathbf{k} = \mathbf{M}^{-1}\mathbf{y}$ by solving the linear system $\mathbf{M} \mathbf{k} = \mathbf{y}$ instead of matrix inversion.

\subsection{Resolving Bifurcations}
\label{sssec:ComposeBifPat}
Given the eigenvectors $\lbrace\mathbf{b}_k\rbrace$ and their respective eigenvalues $\lbrace\lambda_k\rbrace$, we can now compose bifurcation patterns and locate their point of emergence on the trivial branch. Section~\ref{ssec:locating_bif} explains how a bifurcation pattern may be categorised as a simple, multiple or mixed-mode bifurcation. 
For simple bifurcations, each basis vector $\mathbf{b}_k$ defines a bifurcation pattern, which we denote \FixRevThree{as} $\mathbf{x}_b$. We first locate the bifurcation point corresponding to $\mathbf{x}_b$ as follows. Let $\alpha \in \mathbf{p} = \{ {}^uD_u, {}^uD_v, {}^vD_v, {}^vD_u,  {}^uK_u, {}^uK_v, {}^vK_u, {}^vK_v \}$
be the unknown continuation parameter in $\mathbf{p}$. \FixRevOne{Depending on the problem, we use one of the Equations~\ref{eqn:lambda_constraints},~\ref{eqn:cross_exclusive_lambda} or~\ref{eqn:scale_factor_cross} to compute $\alpha$ by plugging in the known parameters from $\mathbf{p}$, and $\lambda_k$ (or $\Lambda = \lambda_k$)}. Then, without loss of generality, we set $u_k = 1$ and solve for $v_k$ using Equation~\ref{eqn:bif_constraints_cross_diff} to compute $\mathbf{x}_b = (u_k\mathbf{b}_k,\,v_k\mathbf{b}_k)$. 
For multiple bifurcations, to compose a bifurcation pattern $\mathbf{x}_b$, \FixRevOne{we first select a set of eigenvectors $\lbrace \mathbf{b}_i\rbrace = \lbrace \mathbf{b}_k\,|\,\lambda_k = \Lambda_m\rbrace$. We then use $\Lambda_m$ as $\lambda_k$ in the case of simple bifurcations to compute $\alpha$ for the corresponding bifurcation point. Next we select an arbitrary set of spectral coefficients $u_i$ to define a desired linear combination of $\mathbf{b}_i$ as an emergent pattern and compute the $v_i$, $\forall i$ using Equation~\ref{eqn:bif_constraints_cross_diff}. Thus we compute the multiple bifurcation pattern $\mathbf{x}_b=(\sum_i{u_i \mathbf{b}_i},\, \sum_i{v_i \mathbf{b}_i})$. 
For a mixed mode bifurcation, 
we pick two sets $\lbrace \mathbf{b}_i \rbrace = \lbrace \mathbf{b}_k \,|\, \lambda_k = \Lambda_m \rbrace$ and $\lbrace \mathbf{b}_j \rbrace = \lbrace \mathbf{b}_k \,|\, \lambda_k = \Lambda_n \rbrace$ and compose an arbitrary linear combination of eigenvectors $\mathbf{b}_k \in \lbrace\mathbf{b}_i \rbrace \cup \lbrace\mathbf{b}_j \rbrace$ to define a desired emergent pattern. We then solve for the bifurcation point $\alpha$ by substituting all parameters from $\mathbf{p}$, and $\Lambda_m$ and $\Lambda_n$ in Equation~\ref{eqn:mixedmode_bifpoint}. With the known bifurcation point and $\Lambda_m$ and $\Lambda_n$, we compute $s_m$ and $s_n$ using Equation~\ref{eqn:mixedmode_sm_sn}. Next, with the previously defined arbitrary values of $u_i$ and $u_j$ for the desired emergent pattern, we compute $v_i = s_m u_i$ and $v_j = s_n u_j$ $\forall i, j$. Finally, we compose the mixed mode bifurcation pattern $\mathbf{x}_b=(\sum_i{u_i \mathbf{x}_i} + \sum_j{u_j \mathbf{b}_j} ,\, \sum_i{v_i \mathbf{b}_i}+ \sum_j{v_j \mathbf{b}_j})$. 
}

\subsection{Reference Method}
\label{ssec:NewtonMethod}
For evaluation purposes, we also implemented a standard approach for detecting bifurcation points~\citep[Chapter $5$]{seydel2010practical}. We call this the \emph{reference method} since it uses common techniques for different tasks in detecting bifurcation points and computing bifurcation patterns. To detect a bifurcation point, we use a test function $\tau$ which is evaluated at ($\mathbf{x}$,$\alpha$) as, 
$\tau = \mathbf{e}_k^T\mathbf{J}\,\mathbf{v}$, 
where vector $\mathbf{v}$ satisfies equation $\mathbf{J}_k\,\mathbf{v} = \mathbf{e}_k$, with $\mathbf{J}_k\, = ( \mathbf{I} - \mathbf{e}_k \mathbf{e}_k^T )\mathbf{J}\, + \mathbf{e}_k \mathbf{e}_k^T$. 
Here, $\mathbf{e}_k$ is a unit vector with all but the $k^{\text{th}}$ element set to zero and $\mathbf{J}$ is the Jacobian matrix for function $\mathbf{f}$ as discussed for branch tracing in Section~\ref{sssec:BranchTracing}. A bifurcation is detected each time $\tau$ changes its sign from $-$ to $+$ in an interval, say $(\alpha_l,\,\alpha_u)$.
Next we perform a \emph{mid-point search} to locate the $0$-crossing for $\tau$ by iteratively evaluating it at $\alpha_{mid} = \rfrac{(\alpha_l + \alpha_u)}{2}$ and then setting $\alpha_l = \alpha_{mid}$ if $\tau < 0$ at $\alpha_{mid}$ or otherwise setting $\alpha_u = \alpha_{mid}$. Upon convergence $|\tau| \approx 0$. This implies that vector $\mathbf{v}$ 
lies in the null-space of the Jacobian matrix $\mathbf{J}$ at $\alpha_{mid}$ and it is a good approximation to the bifurcation pattern $\mathbf{x}_s$. We thus output $\mathbf{x}_s = \mathbf{v}$ and $\alpha_0 = \alpha_{mid}$ as the next detected bifurcation pattern which may be used for branch tracing in a manner similar to the previous methods. Note that this method is not capable of resolving multiple bifurcations.

\subsection{Advantages and Limitations of Proposed Approach}
The main advantage of our proposed approach is that finding bifurcation points and patterns is not dependent on the \emph{goodness} of the tracing test function. 
Using a test function poses two issues: first, the potential presence of more than one bifurcation in an interval, and second, the lack of a guarantee for any test function to detect the presence of each bifurcation. While there are several strategies to address these two issues, often incompletely, our proposed direct approach completely avoids them. Furthermore, we do not need to perform repeated evaluations of the test function along the trivial branch. As a further advantage, depending on the complexity of the test function and the interval size, our approach reduces computational overheads. We tabulate performance gains due to our approach in Section~\ref{sec:experiments}. A third, important  advantage of our approach is that it allows the direct composition of infinite multiple and mixed-mode bifurcation patterns. 
This adds considerable possibilities to bifurcation analysis by supporting the exploration of multiple co-located branches.

At present, the main limitation of our proposed approach is that it can only be used to analyse primary branches emerging from the trivial branch. 
}

\subsection{Branch Tracing \FixRevFive{for Nonlinear Analysis}}
\label{sssec:BranchTracing}
\FixRevFive{Our framework also supports nonlinear analysis with branch tracing in the far-off nonlinear region for the PDEs.}
The first task in branch tracing is to switch over to the new branch of patterns characterised by a given bifurcation pattern $\mathbf{x}_b = (\mathbf{u}_b, \mathbf{v}_b)$. \FixRevOne{While the linear stability analysis gives us $\mathbf{x}_b$, we are interested in a non-trivial solution $\mathbf{x}_s$ that fully satisfies the actual nonlinear PDE given in Equation~\ref{eqn:gen_cross_rd}, near the bifurcation point $\mathbf{p}$ (with $\alpha = \alpha_0$) on the trivial branch. With \dsdkFive{$\mathbf{x}_h = (a_0, b_0)$}\FixRevFive{$\mathbf{x}_h \equiv (a_0, b_0)$} as the homogeneous solution at the bifurcation point $\mathbf{p}$, estimating a good starting point \dsdkFive{(}$\mathbf{x}_s = \mathbf{x}_h + \Delta\mathbf{x} $\dsdkFive{,} \FixRevFive{with }$\alpha \approx \alpha_0$\dsdkFive{)} on the new branch is a non-trivial task. 
The \FixRevFive{nonlinear }solution $\mathbf{x}_s$ must be qualitatively similar to \FixRevFive{the bifurcation pattern }$\mathbf{x}_b$\dsdkFive{ and}\FixRevFive{; }yet far-enough away from the trivial branch to allow continuation without falling back}. 
To achieve this, we propose two improvements over a standard \emph{method of parallel computation} approach for branch switching~\citep[see][Section $5.6.3$]{seydel2010practical}. 

For the first improvement, we suggest using a \emph{bordering algorithm}~\citep{salinger2002loca} to iteratively compute the \emph{jump} \dsdkFive{(}$\Delta\mathbf{x}$\dsdkFive{)} until a successful switch is made. Our key idea is to select a pattern dependent \emph{pivot} (a discrete FEM node) for fixing the jump size and the direction of \emph{parallel computation}. As a second improvement which helps multiple and mixed mode bifurcations we propose to apply a strong guidance to the jump $\Delta\mathbf{x}$ at each intermediate step of our iterative switching algorithm. The bifurcation pattern $\mathbf{x}_b$ serves as a good guidance for the jump $\Delta\mathbf{x}$. The details for our proposed improvements are presented in the supplemental material (SM02.A1). 

Once \dsdkFive{branch switching is accomplished}\FixRevFive{we switch over to a new branch by jumping from the trivial solution $\mathbf{x}_h$ to the nonlinear solution $\mathbf{x}_s$}, we follow \dsdkFive{the new branch}\FixRevFive{it} by means of continuation. Our framework uses the \emph{LOCA} and \emph{NOX} packages from the Trilinos library to perform \emph{pseudo arc-length continuation}. We use an adaptive approach that updates the step size after each continuation step and impose a \emph{tangent scale factor} to manoeuvre the direction of the \emph{continuation curve} as supported by the Trilinos library. In particular, we propose to \dsdkFive{use an}\FixRevFive{establish a} initial tangent direction \FixRevFive{which is strongly orthogonal to the trivial branch} \dsdkFive{based on the orthogonality of $\mathbf{x}_s$ and $\mathbf{x}_h$ for difficult cases}\FixRevFive{by attempting a jump from the nonlinear solution $\mathbf{x}_s$ to its \emph{antithetic} solution $\widehat{\mathbf{x}}_s = \mathbf{x}_h - \Delta\mathbf{x}$, instead of jumping from the trivial solution $\mathbf{x}_h$ to the nonlinear solution $\mathbf{x}_s$}. Again, we present the details of our proposed improvement for branch tracing with other implementation details in the supplemental material (SM02.A1).
\FixRevOne{Also, we provide further details about the configurability of our framework in the supplemental material (SM02.A2)}.

\section{Multi-resolution Adaptation}
\label{sec:adaptation}

\FixRevOne{

Unlike simple spot and stripe patterns, most interesting biological surface patterns exhibit high shape contour irregularities.
The complexity of these patterns is attributed to the surface geometry and the (mid or high) frequency of the constituent eigenmodes (of the LB operator $\nabla^2$). Thus, for accurate computation of such complex patterns, the surface geometry must be represented with a high-resolution FEM discretisation. 
As discussed in Section~\ref{sec:related_work}, many existing geometric and algebraic multigrid approaches support high-resolution meshes but face challenges in either performance, scalability or parallelisability for branch tracing owing to tightly coupled multiple levels.  
We thus propose a simple multi-level approach in which branch tracing is performed at the base-level of discretisation and the resultant patterns are upsampled and resolved at higher levels progressively.                 
}

{\color{AddedNew3}
\begin{wrapfigure}[7]{r}{0.4\textwidth}
\vspace{-0.65cm}
\includegraphics[width=0.39\textwidth]{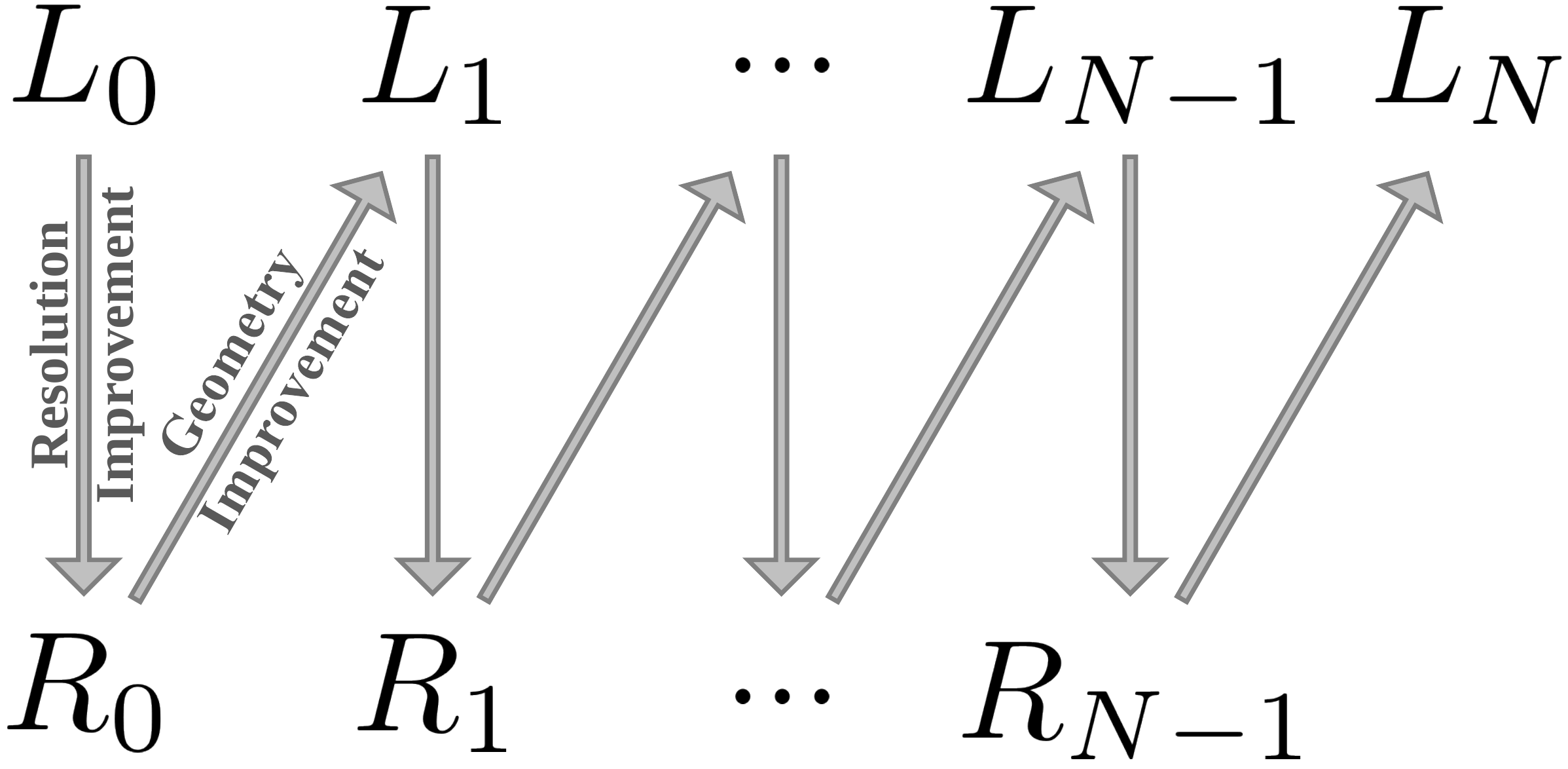}
\end{wrapfigure}
Our framework uses a simplified geometric multigrid approach where the surface domain $\Omega$ is organized in multiple levels as \FixRevTwo{$L_l$, with $l= 0, \ldots,  N$, where $L_0$ is the lowest resolution representation and $L_N = \Omega$}. 
We begin with the highest level mesh $L_N = \Omega$ and generate each lower level mesh $L_{l-1}$ from mesh $L_{l}$ by \FixRevOne{applying a \emph{quadric-based edge collapse decimation} algorithm~\citep{garland1997surface,cignoni2008meshlab}}.
We perform branch tracing at the lowest resolution  $L_0$ and progressively upsample resulting patterns up to the highest resolution $L_N$. Unlike most of the multigrid approaches where all the mesh levels are used simultaneously in a \emph{V-cycle} or a \emph{W-cycle} (for the full multigrid approach) we perform complete upsampling of a solution using only two levels at a time. We thus call our approach a \emph{progressive geometric multigrid} approach.

We propose a two-step approach for upsampling the results between two levels $l-1$ and $l$. For the first step, we increase the mesh resolution for $L_{l-1}$ without changing its geometry. We perform an in-plane subdivision of each existing (quad) finite element into four to give a new mesh, say $R_{l-1}$. In the second step, we map the geometry of the mesh $R_{l-1}$ to $L_{l}$. For each solution vector $\mathbf{x}_b^{l-1}$ defined over $L_{l-1}$ with continuation parameter $\alpha = \alpha_b$, we first interpolate it linearly to the higher resolution mesh $R_{l-1}$ and then solve the nonlinear system $\mathbf{f}(\mathbf{r}_b^{l-1}, \alpha_b) = 0$ over $R_{l-1}$. In general, we use a Newton method with \emph{backtracking} to solve for $\mathbf{r}_b^{l-1}$. We compute the direction vector for the Newton method using a \emph{biconjugate gradient method with stabilization} as implemented in the AztecOO package. For difficult cases, we use a \emph{trust region method} for solving the above nonlinear system with a \emph{GMRES} approach for establishing the search direction.    Next we perform a similar interpolation from $R_{l-1}$ to $L_{l}$ and then solve for $\mathbf{x}_b^{l}$\FixRevThree{ with $\mathbf{f}(\mathbf{x}_b^{l}, \alpha_b) = 0$}. 

For linear interpolation of a solution across two meshes (say, from $L_{l-1}$ to  $R_{l-1}$  or from $R_{l-1}$ to $L_{l}$) we use \FixRevOne{a projection based mapping scheme between meshes. We project each node from a target mesh (say $R_{l-1}$) onto the nearest face of the source mesh ($L_{l-1}$) and use the barycentric coordinates of the projected node to compute linear interpolation weights. 
A naive implementation of this mapping scheme has a computational complexity $\mathcal{O}(MN)$ with $M$ and $N$ as the number of nodes for the source and target meshes respectively. Computational complexity can be reduced to $\mathcal{O}(M\log{}N)$ with the use of a \emph{kd-tree} for the nearest face search. Note that we compute the multilevel meshes and the mappings for linear interpolation of solutions only once as a preprocessing step for a given surface domain $\Omega$.}

Our two-step approach separates the complexity \dsdk{due to a higher resolution from the complexity due to geometric changes in surface representation}\FixRevThree{of resolution improvements from the complexity of geometry improvements} for an upsampling task. 
The results for our progressive geometric multigrid approach are presented in Section~\ref{sec:experiments}.

\paragraph{Parallelisability.}
\label{sec:parallel}
{
Branch tracing is inherently a sequential task, because the next solution on a branch is computed by numerically analysing the previous one. For large-scale systems, \citet{salinger2005bifurcation,lehoucq2001large} suggest using a multi-processor environment where a large domain ($\Omega$) is partitioned into sub-domains and each domain is solved on a separate processor. Their domain-partitioning approach is generic but the parallelism is limited to solving for one solution at a time while tracing the branch sequentially. Similarly, Continillo ~\citep{continillo2012parallel} present and discuss a method for parallelisation of the most repeated operation for their reactive dynamical system, i.e. the computation of the \emph{Jacobian Matrix} $\mathbf{J}$. They parallelise the computation of $\mathbf{J}$ using a \emph{cluster} with a \emph{message passing interface (MPI)}. Again their approach limits parallelisation to the computation of the next solution. 
}

Our framework has considerable potential for parallelism. The most important of it is the scope for \dsdk{solving}\FixRevThree{upsampling} several patterns on a branch simultaneously. This is a direct outcome of our progressive geometric multigrid approach which completely delineates the branch tracing problem from the convergence of a solution at a higher resolution. Once a branch is traced at the lowest resolution with domain $\Omega_0$, our framework can be invoked to resolve each of the patterns on the branch independently. This allows several instances of our framework to be launched on a computer cluster to simultaneously solve for all the solutions on a branch at higher resolutions. We evaluate the overhead of such parallel launches due to file reading and mesh data preparation operations in Section~\ref{sec:experiments} and provide theoretical estimates for potential gains from parallel computation of a given branch. 
}

\section{Experiments and Results}
\label{sec:experiments}
In this section, we describe our experimental setup, discuss case studies and present results. We use a $64$-bit \emph{Ubuntu $14.04$ LTS} platform running on an \emph{Intel Xeon E5-2630} CPU with $6$ Cores @$2.3$Ghz with $16$GB RAM for all performance evaluations and most other experiments. We demonstrate our framework with two RD system case studies, a Brusselator system for the cotyledon patterning of conifer embryos (Section~\ref{sssec:brusselator_adaption}) and Murray's chemotactic model for snake-skin pattern formation (Section~\ref{sssec:murray_adaption}). 

\subsection{Case Study I: A Brusselator model}
\label{sssec:brusselator_adaption}

\citet{nagata2013reaction} present a study of emergent cotyledon patterns on a plant tip. They model observed patterns as bifurcations for a two-component Brusselator RD system. They represent the plant tip with a \FixRevOne{simple} geometric shape, a \emph{spherical cap}. This spherical cap domain $\Omega$ is parameterised by its size 
\FixRevOne{
factor 
}
$R$ and a curvature factor $\zeta$ 
\FixRevOne{
as shown in Figure~\ref{fig:spherical_cap}. 
} 
\citet{nagata2013reaction} perform marginal stability analysis to 
\FixRevOne{
study the relation between the number of emergent cotyledons and the size factor $R$ or the curvature factor $\zeta$ for a cap.}
Mathematically, their model is defined as,
\begin{align}
\frac{\partial a}{\partial t}\,\,&=\,\, D_1 \nabla^2 a \,+\, A \, - \,(D + B) a\,  + \,C a^2 b\,, \quad\quad\quad
\frac{\partial b}{\partial t}\,\,=\,\, D_2 \nabla^2 b \,+\, B a \, - \,C a^2 b\,, \nonumber\\
&\text{with b.c.}\quad a(\mathrm{d}\Gamma) = \frac{A}{D},\quad b(\mathrm{d}\Gamma) = \frac{B D}{A C},\quad \mathrm{d}\Gamma \in \Gamma\,.
\label{eqn:bruss}
\end{align} 
\begin{wrapfigure}[12]{r}{0.28\textwidth}
\vspace{-0.7cm}
\begin{tikzpicture}
\node[anchor=south west,inner sep=0] (image) at (0,0) {
\includegraphics[width=0.25\textwidth]{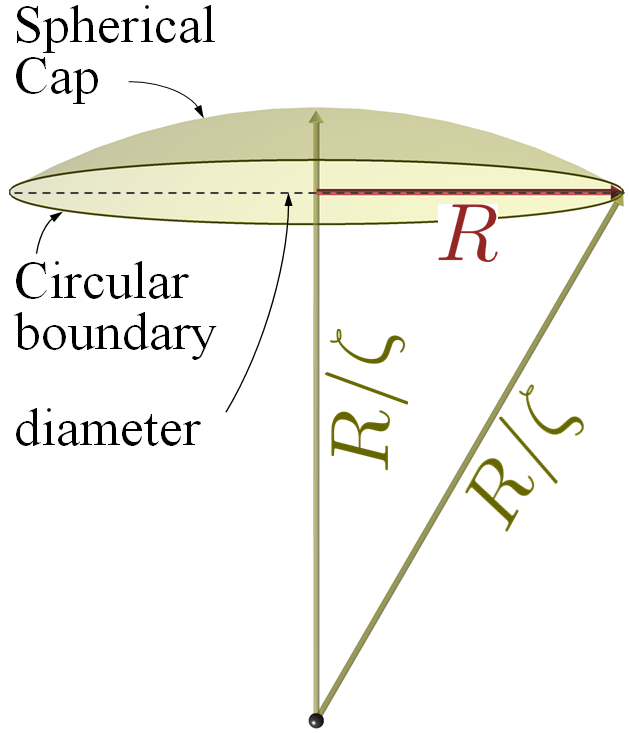}
};
\begin{scope}[x={(image.south east)},y={(image.north west)}]
    \end{scope}
\end{tikzpicture}    
\caption{A spherical cap.}
\label{fig:spherical_cap}
\end{wrapfigure}
Here, 
\FixRevOne{
$a$ and $b$ are concentrations of two Turing morphogens}%
, $A$, $B$, $C$ and $D$ are positive rate constants, $D_1$ and $D_2$ are positive diffusion rates and $\mathrm{d}\Gamma$ is an infinitesimal element on the boundary $\Gamma$ for the domain $\Omega$. 
\FixRevOne{
Nagata et al. linearise Equation~\ref{eqn:bruss} near homogeneity with $(a_0 = \rfrac{A}{D},\, b_0 = \rfrac{B D}{A C})$ to derive an analytical expression for their continuation parameter (represented by $A$ here) at a \emph{simple} bifurcation point. \mze{The continuation parameter $A$ is determined} by the other system parameters and the eigenvalue $\Lambda = \lambda_i$ of a spherical \mze{cap} harmonic $\phi_i$, which constitutes the emergent pattern. \mze{Note that $\lambda_i$, in turn, depends on the shape $\zeta$ and size $R$ of the cap, and it can be computed by evaluating an \emph{associated Legendre function}}. This makes it possible to study the \emph{marginal stability} of emergent patterns with respect to $A$, $R$ and $\zeta$.
\mzk{Finally, note that said linearisation results in zero \emph{Dirichlet} boundary conditions which we simulate in our framework as well.} }

\FixRevOne{
In our framework, the spherical cap harmonics are just a special case of eigenfunctions $\lbrace\phi_i\rbrace$ for a specific, simple domain. The benefit of our approach, however, is that we can easily generalise the analysis to arbitrarily shaped domains. This could facilitate the discovery of shape-induced anomalies in emergent patterns, as we discuss in an example later. \mze{Further, we simplify the analysis by incorporating the size factor $R$ directly into the RD model. We replace the size $R$ of the spherical cap with a factor $\gamma$ that represents the relative scale of an arbitrary domain with respect to its canonic unit size.} 
We then numerically compute the emergent patterns and corresponding $\Lambda$ values. }
   
We first modify Equation~\ref{eqn:bruss} to include the scale factor $\gamma$ in all the reaction terms, that is, \FixRevOne{in all coefficients except the diffusion rates $D_1$ and $D_2$. Scaling the reaction relative to the diffusion coefficients has the same effect as scaling the domain~\citep[page $78$]{murray2003mathematical}. We use the notation $A = \gamma A_\ast$, where $\ast$ indicates corresponding rates at unit scale, and similarly for the other coefficients.} This yields new linearisation parameters
\begin{align}
\quad {}^uK_u = \gamma\,(B_* - D_*),\quad {}^uK_v = \gamma\,\frac{A_*^2C_*}{D_*^2}&,\quad {}^vK_u = - \gamma\,B_*,\quad {}^vK_v =   - \gamma\,\frac{A_*^2C_*}{D_*^2}\,.  
\label{eqn:bruss_params_with_scale}
\end{align}   
\FixRevOne{Substituting these parameters along with ${}^uD_u=D_1$, ${}^uD_v=0$, ${}^vD_v=D_2$, ${}^vD_u=0$ and $\lambda_i=\Lambda$ in Equation~\ref{eqn:lambda_constraints} gives us} the new continuation parameter $A_*$ as
\begin{align}
A_*\,\,&=\,\, D_* \left(\frac{ D_2 ( B_* - D_*) \Lambda_* \,-\,D_1 D_2 \Lambda_*^2}
{C_* \left( D_1 \Lambda_*\,+\, D_* \right)}\right)^{\rfrac{1}{2}} \,,\,
\text{with}\quad \Lambda_* \,\,=\,\,\frac{\Lambda}{\gamma}\,\,.
\label{eqn:bruss_params_A*}
\end{align}
\FixRevOne{Here $\Lambda$ denotes the eigenvalue for a pattern $\phi_i$ at unit scale of the domain $\Omega$. It can be shown that under uniform scaling of a surface domain $\Omega$ by a factor $R$, the eigenvalue for a given eigenfunction is inversely proportional to $R^2$. For spherical cap harmonics, for example, eigenvalues are analytically defined as $\lambda_i = \rfrac{\Psi(\phi_i,\,\zeta)}{R^2}$, where $\Psi$ is defined in terms of an \emph{associated Legendre function} dependent on the wavemode $\phi_i$ and \FixRevFive{the curvature factor }$\zeta$ \citep[see][Equation $11$]{nagata2013reaction}. Thus substituting $\gamma=R^2$ in our modified Brusselator model allows us to incorporate the size factor $R$ for studying marginal stability with arbitrary domains. 
}

We compute each eigenmode $\phi_i$ and its respective eigenvalue $\Lambda$ numerically only for a representative arbitrary shape $\Omega$ \FixRevOne{at a unit scale. 
 Our modified Brusselator model then} enables us to plot marginal stability curves for different bifurcation patterns on \FixRevOne{this} arbitrarily shaped plant tip against \FixRevOne{the size factor $R$}, similar to the plots for spherical caps as presented by \citet{nagata2013reaction} in Figure $3$ of their paper. 
\subsubsection*{Experiments.} 
\begin{figure*}[t]
\begin{tabular}{ccccccc}
\includegraphics[width=0.11\linewidth]{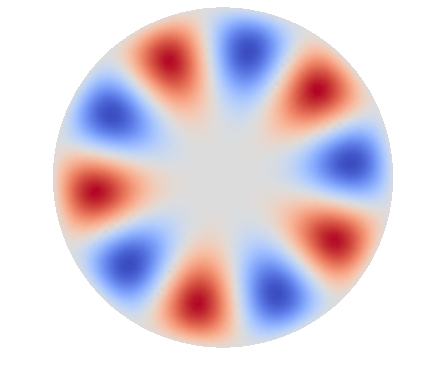} &
\includegraphics[width=0.11\linewidth]{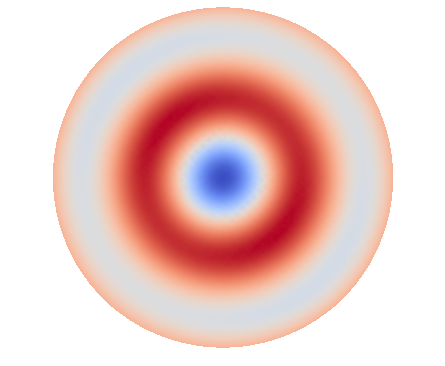} &
\includegraphics[width=0.11\linewidth]{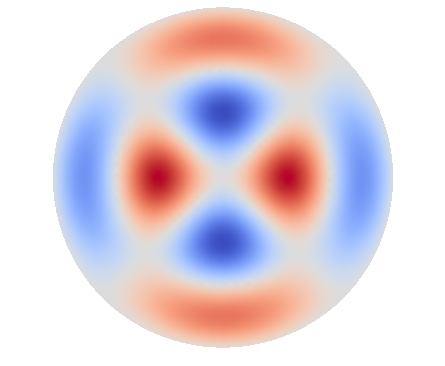} &
\includegraphics[width=0.11\linewidth]{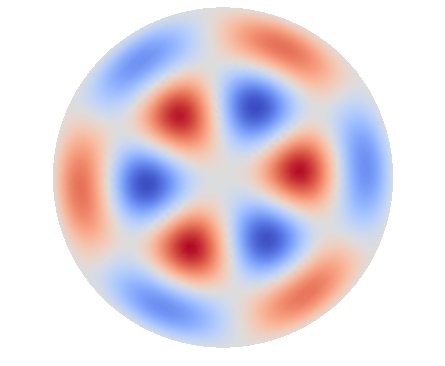} &
\includegraphics[width=0.11\linewidth]{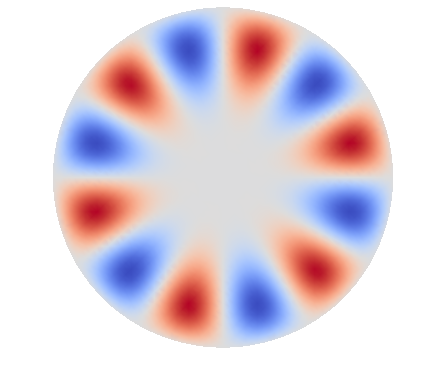} & 
\includegraphics[width=0.11\linewidth]{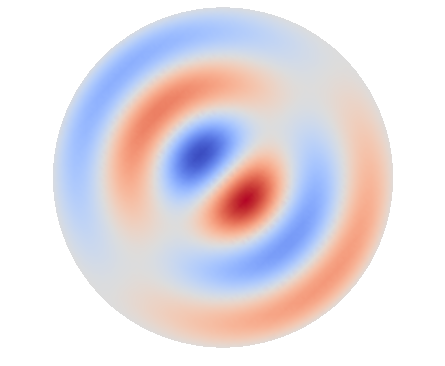} & 
\includegraphics[width=0.11\linewidth]{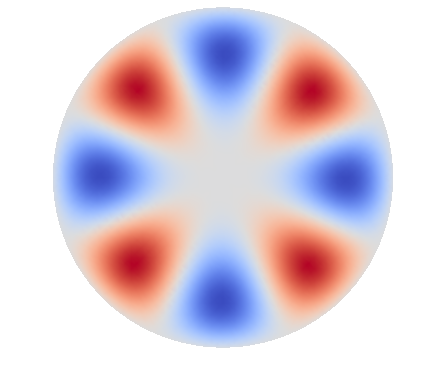} \\
$(5,\,1)$ &
$(0,\,3)$ &
$(2,\,2)$ &
$(3,\,2)$ &
$(6,\,1)$ &
$(1,\,3)$ &
$(4,\,1)$ 
\end{tabular}
\caption{Spherical cap harmonics in order of their emergence along the trivial branch with $(a_0 = \rfrac{A_*}{D_*},\,b_0 = \rfrac{B_* D_*}{A_* C_*})$ \mze{for a spherical cap}.}
\label{fig:bruss_bif_patterns}
\end{figure*}
{
\setlength{\extrarowheight}{.5ex}
\begin{table*}[t]
\scriptsize
  \centering
  \caption{Errors in locating bifurcation points \FixRevOne{$A_*$} for the Brusselator model acting on a spherical cap domain. We compare the \emph{proposed} and \emph{reference} methods vis-a-vis \FixRevOne{analytically defined results} \FixRevFive{with lookup tables from~\citet{bauer1986tables}, }for several emergent modes. \MZ{say 'bifurcation location A*' instead of 'method'; make order of columns consistent; call error columns 'relative error', then 'Proposed' and 'Reference'}}
\begin{tikzpicture}
\node[anchor=south west,inner sep=0] (image) at (0,0) {
    \begin{tabulary}{\textwidth}{|C|ccc|CC|}
    \hline
    \textbf{Mode} & \multicolumn{3}{c|}{\textbf{Bifurcation Point Location ($ A_*$)}}   & \multicolumn{2}{c|}{\multirow{1}[0]{*}{\textbf{Relative Error}}} \\
    \cline{2-4}\cline{5-6}
    \multirow{2}[0]{*}{\textbf{$\Phi_i$ : (m, n)}} & \textbf{Analytical} & \textbf{Reference} & \textbf{Proposed } 
&\multicolumn{1}{c}{ RE0}&  \multicolumn{1}{c|}{ RE1} \\   
            & \textbf{A0} & \textbf{A} & \textbf{A1} & \textbf{(A - A0) / A0} & \textbf{(A1 - A0) / A0} 
          \\
          [\dimexpr-\normalbaselineskip+3\smallskipamount] \hline
    (5, 1)  & 0.76520 & 0.76528 & 0.76528 & 1.07E-04 & 1.07E-04 \\
    (0, 3)  & 0.76382 & 0.76403 & 0.76403 & 2.77E-04 & 2.76E-04 \\
    (2, 2)  & 0.76171 & 0.76200 & 0.76199 & 3.84E-04 & 3.74E-04 \\
    (3, 2)  & 0.76049 & 0.75997 & 0.75997 & -6.88E-04 & -6.91E-04 \\
    (6, 1)  & 0.75552 & 0.75475 & 0.75475 & -1.02E-03 & -1.02E-03 \\
    (1, 3)  & 0.75406 & 0.75343 & 0.75343 & -8.41E-04 & -8.37E-04 \\
    (4, 1)  & 0.74744 & 0.74792 & 0.74793 & 6.41E-04 & 6.49E-04 \\

	\hline    
    \end{tabulary}%
};
\begin{scope}[x={(image.south east)},y={(image.north west)}]
    \end{scope}
\end{tikzpicture}        
  \label{tab:cap_bif_locations}%
\end{table*}%
}

First we validate our framework \FixRevOne{against analytically derived} results from \citet{nagata2013reaction}. We use the same parameter values as given by Nagata et al. for their Figure $3$ (i.e. $D_1 = 0.005$, $D_2 = 0.1$, $B_* = 1.5$, $C_* = 1.8$, $D_* = 0.375$) and a spherical cap with radius $R=1$ and curvature factor $\zeta = 0.5$. 
\FixRevOne{For all quantitative evaluations in this section,} we compute seven different emergent patterns \FixRevOne{as shown in Figure~\ref{fig:bruss_bif_patterns}} and their corresponding bifurcation points $A_\ast$ using both our \emph{proposed method}, and the \emph{reference method}, as explained 
in Section~\ref{sec:framework}. 
\FixRevOne{We begin by examining errors in computing $A_\ast$ with an FEM mesh of order one with about $4300$ nodes.}
\begin{figure}[t]
\centering
\begin{tikzpicture}
\node[anchor=south west,inner sep=0] (image) at (0,0) {
\includegraphics[width=0.8\textwidth]{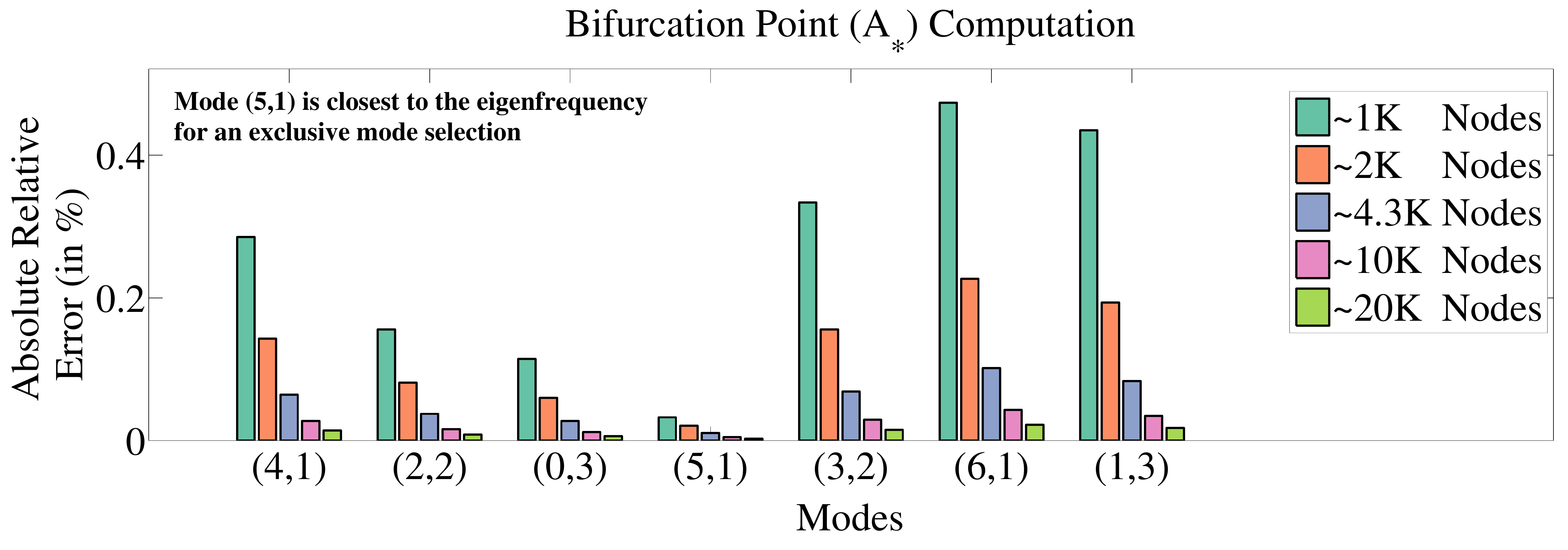}
};
\begin{scope}[x={(image.south east)},y={(image.north west)}]
    \end{scope}
\end{tikzpicture} 
\caption{Error in computing bifurcation points $A_*$ for seven emergent wavemodes $\phi_i$ using different mesh resolutions. The plot shows relative errors in comparison with analytically derived values for different wavemodes in order of their corresponding eigenvalues. \MZ{show absolute relative error}}
\label{fig:bruss_point_errors}
\end{figure}
Table~\ref{tab:cap_bif_locations} provides a numerical comparison of relative errors \FixRevOne{for both methods} against the \FixRevOne{analytically derived} values. 
For all emergent patterns, 
our proposed method has a low relative error \FixRevOne{on the order of $10^{-3}$} when compared with the expected analytical results, and \FixRevOne{the errors are quantitatively similar to those for the reference method. }
This implies that our direct approach works well and most of the errors can be explained as approximation errors due to FEM discretisation.

\MZ{explain order of eigenmodes in the figure; explain why error is lowest at exclusive mode} Figure~\ref{fig:bruss_point_errors} shows relative errors for locating all seven bifurcation points 
at five mesh resolutions ranging from $1K$ - $20K$ vertices. The bifurcation points are ordered according to increasing eigenvalues along the horizontal axis. We found the relative error to fall quickly as a power-law function of the mesh resolution for each pattern. 
\dsdk{We observe that the first emergent mode $(5,1)$ along the trivial branch has lowest error magnitudes, and we expect the \emph{high-frequency} patterns corresponding to large eigenvalues to have larger relative errors in locating bifurcation points.} 
\FixRevTwo{Mode $(5,1)$ is closest to the exclusive mode selection criterion in Equation~\ref{eqn:cross_exclusive_lambda} and it has the lowest relative error. We observe that the errors increase with the distance from this reference mode $(5,1)$ in the eigenspectrum and expect larger relative errors in locating bifurcation points for  \emph{higher frequency} patterns}.
Thus for applications requiring high accuracy for studying emergent patterns with arbitrary domains, \FixRevOne{our multi-resolution approach presented in Section~\ref{sec:adaptation} \mzk{and supporting meshes with FEM nodes on the order of $10^6$}is beneficial. }       

Next, we examine numerical errors in computing bifurcation patterns for our proposed method in comparison with analytically defined spherical cap harmonics (i.e. the ground truths).  
Figure~\ref{fig:bruss_pattern_errors_diff_resolution} shows relative root mean square (RMS) errors in 
emergent patterns with different mesh resolutions. 
\FixRevOne{For comparison we normalize all spherical cap harmonics to \FixRevThree{have} unit amplitude}. Again we see that the accuracy improves with the discretisation resolution with some power-law function. We also include the errors for the reference method for a mesh with $4300$ nodes \FixRevThree{in Figure~\ref{fig:bruss_pattern_errors_diff_resolution}}, which indicates that our proposed method is quantitatively consistent with the reference method. 

We also studied the impact of the FEM \emph{order} on RMS errors with about $5000$ nodes in each case, see Figure~\ref{fig:bruss_pattern_errors_diff_fem_order}. We denote surface geometries that are approximated with finite elements of order $1$, $2$ and $3$, each with about $5000$ nodes, by $G1$, $G2$, and $G3$. 
\FixRevOne{Since FEM of increasing order has an increasing number of nodes per \emph{planar} finite element, the spherical surface geometry is approximated with a decreasing number of planar elements for higher orders. Comparing the solutions of order $1$ FEM on $G1$, order $2$ on $G2$, and order $3$ on $G3$, we see that FEM order $1$ outperforms higher orders for all emergent patterns. At the given resolution of $5K$ FEM nodes, the increase in error due to the geometric approximation in $G2$ and $G3$ \FixRevOne{apparently} outweighs the error reduction due to higher order polynomials. In addition, we also report the error of order $1$ FEM on the lower resolution geometries $G2$ and $G3$. As expected this increases the error due to the use of lower order polynomials, but only marginally.
}
\begin{figure}[t]
\centering
\begin{subfigure}[t]{0.49\linewidth}
\centering
\begin{tikzpicture}
\node[anchor=south west,inner sep=0] (image) at (0,0) {
\includegraphics[height=0.54\linewidth]{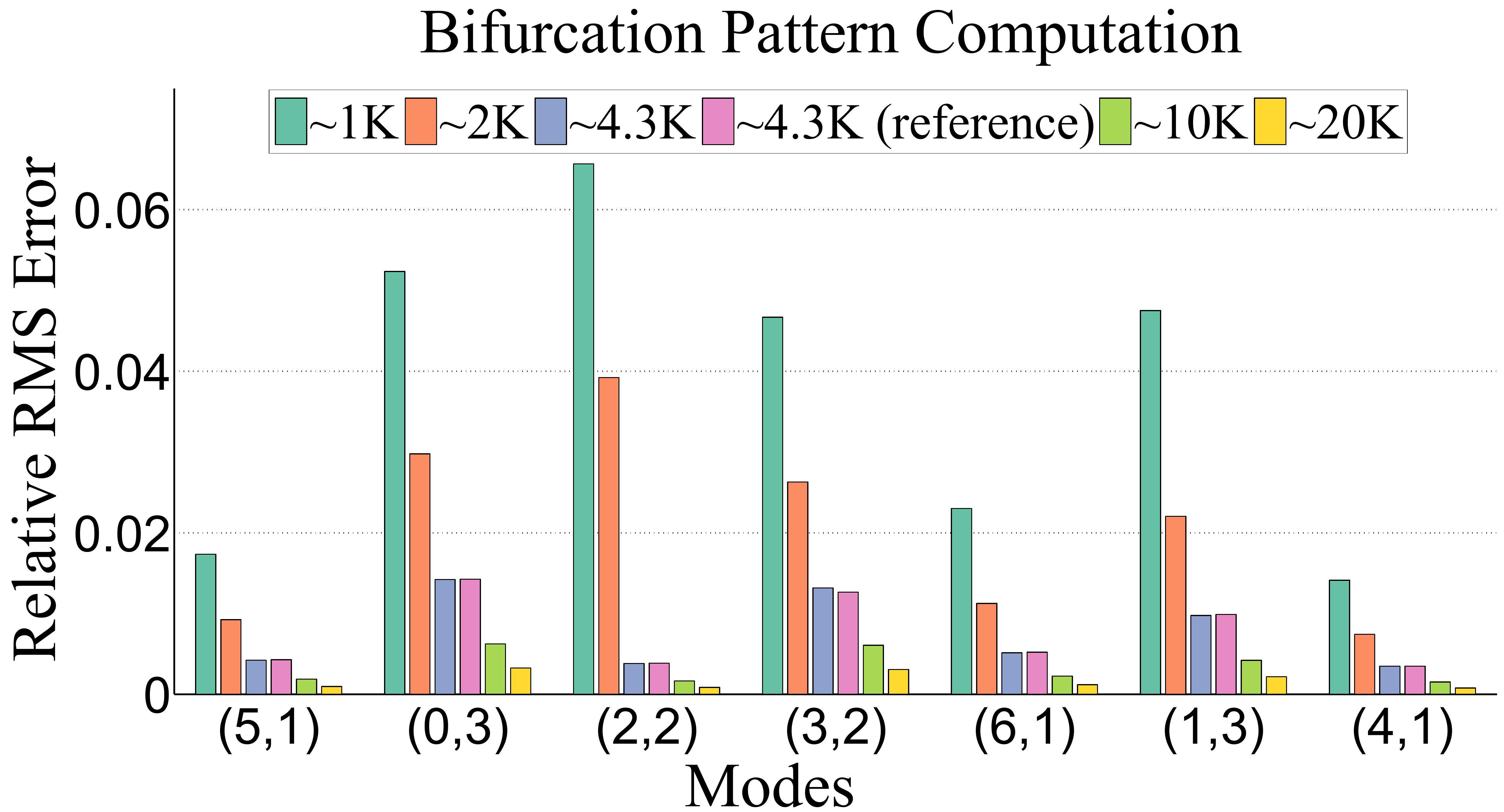}
};
\begin{scope}[x={(image.south east)},y={(image.north west)}]
    \end{scope}
\end{tikzpicture} 
\caption{Different resolutions}
\label{fig:bruss_pattern_errors_diff_resolution}
\end{subfigure} \hfill{}
\begin{subfigure}[t]{0.49\linewidth}
\centering
\begin{tikzpicture}
\node[anchor=south west,inner sep=0] (image) at (0,0) {
\includegraphics[height=0.54\linewidth]{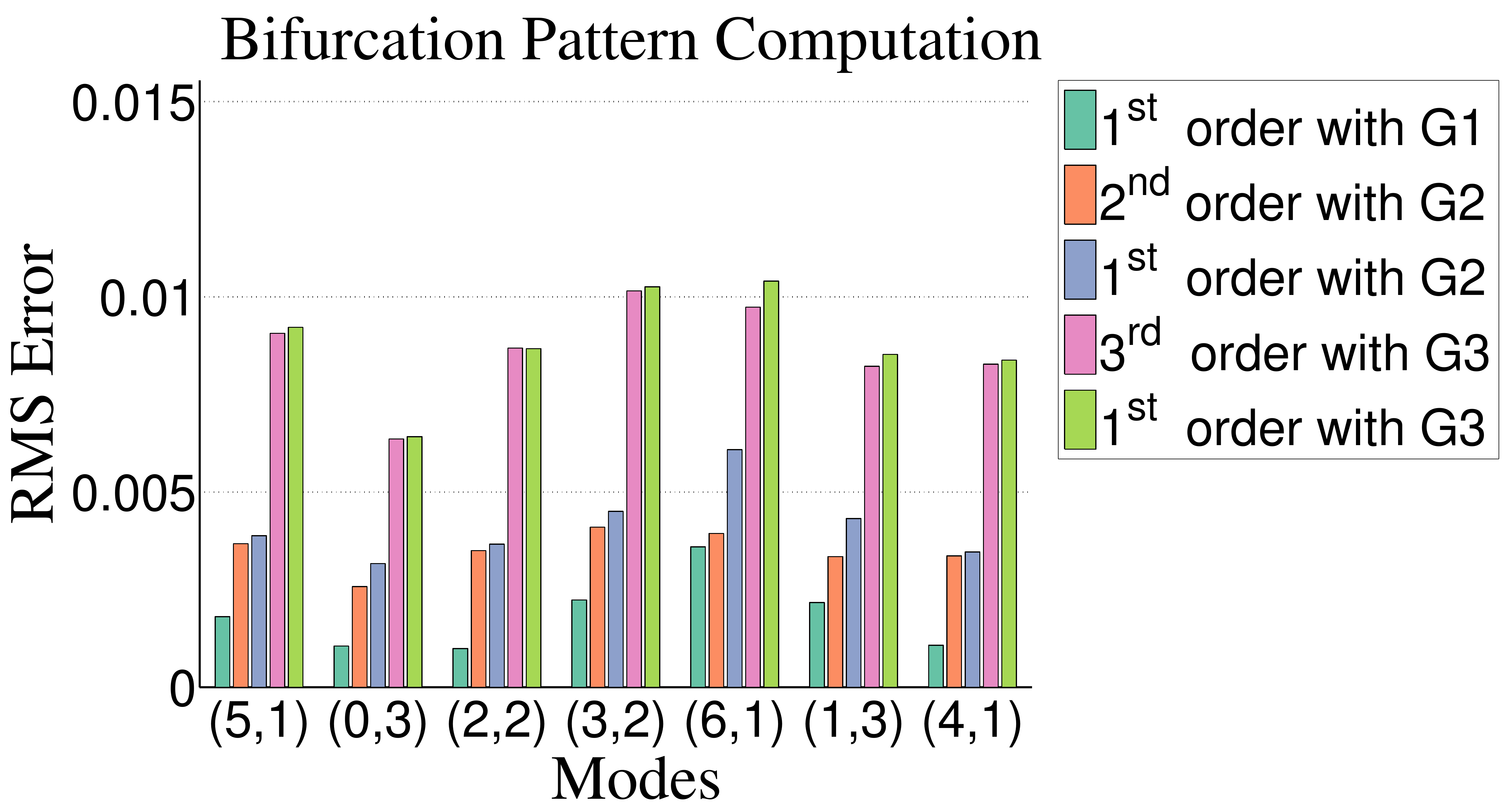}
};
\begin{scope}[x={(image.south east)},y={(image.north west)}]
    \end{scope}
\end{tikzpicture} 
\caption{Different FEM order with $~5K$ mesh}
\label{fig:bruss_pattern_errors_diff_fem_order}
\end{subfigure}
\caption{Relative mean error in computing the wavemode pattern $\phi_i$.}
\end{figure}

\FixRevOne{Next, we illustrate the key advantage of our approach, that is, the ability to study the effects of arbitrary shape distortions on emergent patterns. Figure~\ref{fig:bruss_distortion_patterns} shows one such distorted shape in its second row, which we generate by deforming the circular boundary of the cap and propagating the distortions smoothly over the entire surface.
We visualize the effects of these shape distortions on \FixRevOne{three }emergent patterns using a nonlinear color mapping, as illustrated in the figure. Clearly, new patterns show marked deviations from the respective spherical cap harmonics shown at the top row. Deviations \FixRevOne{from} normal emergent patterns often lead to developmental anomalies, which are studied extensively in mathematical biology, refer to \citet{harrison2004spatially} for an example. While our illustration here does not explain any specific anomaly, our framework can certainly be used to study the role of domain shape deviations on actual observed cases. 
}
\begin{figure}
\centering
\begin{tikzpicture}
\node[anchor=south west,inner sep=0] (image) at (0,0) {
\includegraphics[width=0.7\linewidth]{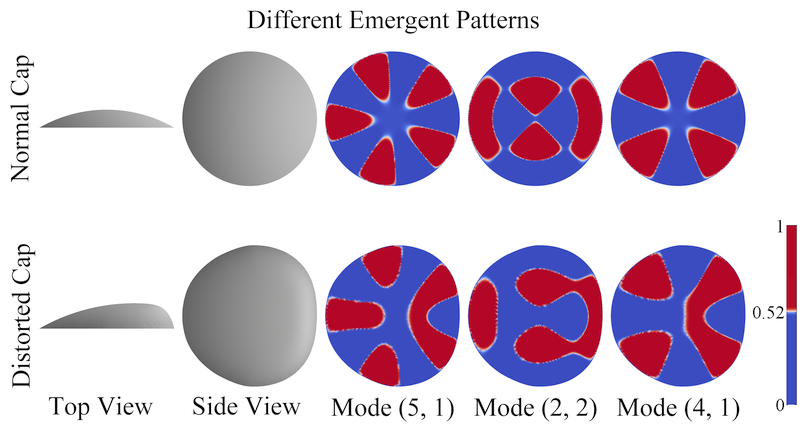}
};
\begin{scope}[x={(image.south east)},y={(image.north west)}]
    \end{scope}
\end{tikzpicture} 
\caption{Effects of shape distortion on emergent patterns.}
\label{fig:bruss_distortion_patterns}
\end{figure}

Finally, we present marginal stability curves for mode $(5,1)$ in Figure~\ref{fig:bruss_marginal_stability}. Here we consider three cases: (a) a spherical cap growing isotropically (\FixRevOne{greenish} colour), (b) a distorted cap growing isotropically (\FixRevOne{brownish} colour), and (c) a spherical cap that first grows isotropically until $R=1$ and then progressively distorts with further growth (\FixRevOne{blue-gray} colour). 
We first computed eigenvalues $\Lambda$ for the spherical cap with $R=1$, corresponding to a boundary perimeter $P=2\pi R$, and $\zeta = 0.5$, and  for the distorted cap with boundary perimeter $P = 2\pi$. Using these eigenvalues we plot solid curves in Figure~\ref{fig:bruss_marginal_stability} for $A_*$--vs.--$R$ (or $P$)  with $\gamma = \rfrac{P^2}{4\pi^2}$ for Equation~\ref{eqn:bruss_params_A*}. Our numerically computed marginal stability curve for the isotropically growing spherical cap domain conforms well with expected theoretical values \FixRevOne{shown using greenish circles}~\citep{nagata2013reaction}. The brown solid curve in Figure~\ref{fig:bruss_marginal_stability} shows that our framework can perform a similar study for an arbitrary shape domain. \FixRevOne{Last but not least, we demonstrate our capability to study marginal stability with a dynamically changing arbitrary shape. We progressively morph the spherical cap into a  distorted cap as in case (b) with a shape-blending factor $\alpha$ that varies from $0$ to $1$ for $P= 1\times 2\pi$ to $P= 1.2\times 2\pi$. The blue-gray solid curve in Figure~\ref{fig:bruss_marginal_stability} shows the respective stability curve.} 

\begin{figure*}[t]
\centering
\begin{tikzpicture}
\node[anchor=south west,inner sep=0] (image) at (0,0) {
\includegraphics[width=\linewidth]{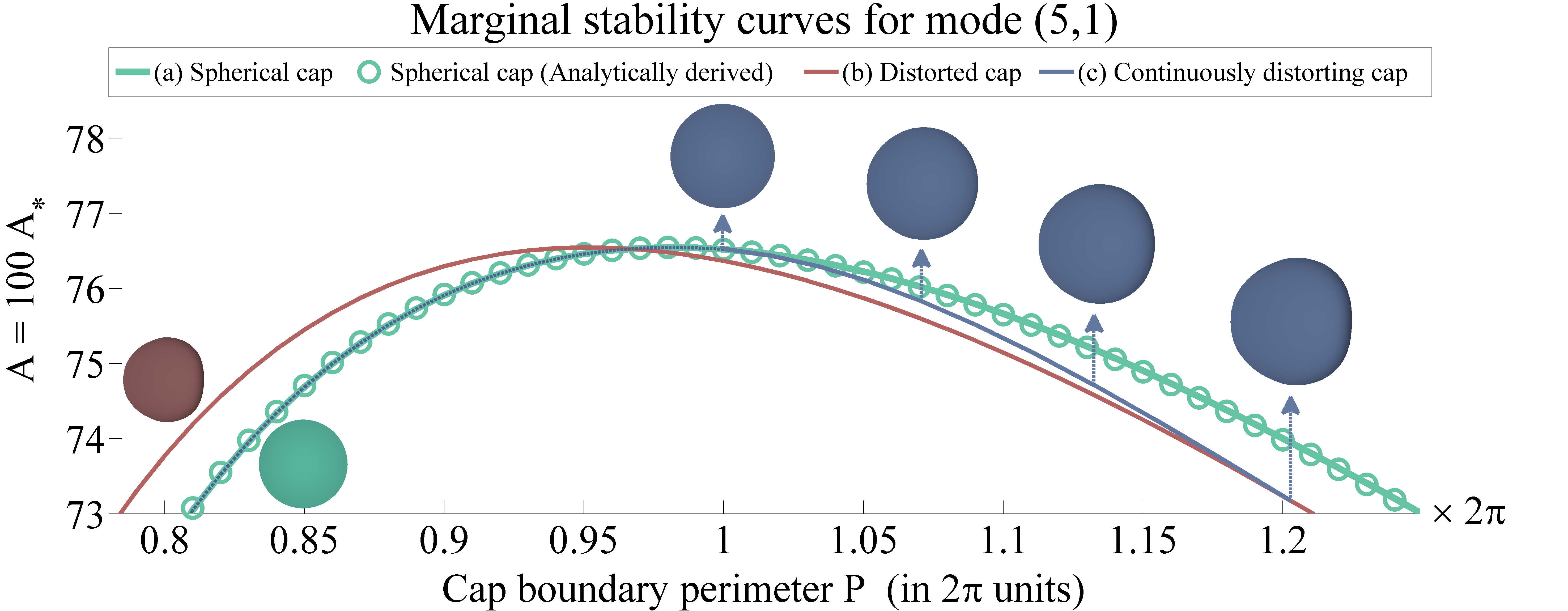}
};
\begin{scope}[x={(image.south east)},y={(image.north west)}]
    \end{scope}
\end{tikzpicture} 
\caption{Marginal stability for mode $(5,1)$ in $A_*$\emph{ -vs- }$P$ (cap boundary perimeter in $2\pi$ units) parameter slice. {Analytically derived} values as in Figure $3$ by \citet{nagata2013reaction} for an isotropically growing spherical cap are indicated with circular markers. Solid lines indicate results for our proposed method.}
\label{fig:bruss_marginal_stability}
\end{figure*}

\subsection{Case Study II: Murray's model}
\label{sssec:murray_adaption}
\citet{winters1990tracking} present bifurcation analysis of a two-component RD system with nonlinear reaction terms to model snakeskin pigmentation. They consider the role of \emph{chemotaxis}, which is expressed in terms of surface gradients of the components. The nondimensional form of their system is given as
\begin{align}
\frac{\partial a}{\partial t} \, &= \, D \nabla^2 a \,-\,\alpha \nabla \cdot \left(a \nabla b\right) \,+\, S \,C\, a (N - a)\,,\quad\quad\quad
\frac{\partial b}{\partial t} \, = \, \nabla^2 b \,+\,S\left( \frac{a}{1+a} \,-\,b\right)\,, \nonumber\\
&\text{with b.c. }\quad \widehat{n}\cdot \nabla b (\mathrm{d}\Gamma)\,=\,  
\widehat{n}\cdot \nabla a (\mathrm{d}\Gamma)\,=\,0\,,\quad
\mathrm{d}\Gamma \in \Gamma\,.  
\label{eqn:murray_nonlin}
\end{align}   
Here, $a$ represents the \emph{melanophore} cell density, $b$ the concentration of a \emph{chemoattractant} that attracts the melanophores, $D$ is the cell diffusion rate within the \emph{cell matrix} over the surface, $\alpha$ is the strength of chemoattraction, $C$ represents the \emph{cell mitotic rate}, $S$ is a \dsdk{relative}\FixRevThree{positive} scale factor\dsdk{for the domain $\Omega$}, and $N$ is the maximum cell concentration capacity for the growth model. 
All these parameters are positive and uniform over the entire snakeskin. Finally, $\mathrm{d}\Gamma$ is an infinitesimal element on the boundary $\Gamma$ of $\Omega$, $\widehat{n}$ is the outward surface normal at $\mathrm{d}\Gamma$, and the model uses zero Neumann boundary conditions.  

\dsdkFive{Winter et al.}\FixRevFive{\citet{winters1990tracking}} perform bifurcation detection and branch tracing to discover the steady-state solutions of the system in Equation~\ref{eqn:murray_nonlin} for $2D$ rectangular domains.
They employ a second-order FEM approximation to represent the system state with a vector $\mathbf{x} = (a,\,b)$, and a standard \emph{Galerkin weak-formulation} for a discretised representation of temporal derivatives given in Equation~\ref{eqn:murray_nonlin} with $\mathbf{f}(\mathbf{x}, \mathbf{p}, \alpha) = (\rfrac{\partial a}{\partial t},\,\rfrac{\partial b}{\partial t})$. Here, $\mathbf{p} = \left\lbrace C,\,D,\,N,\,S \right\rbrace$ is the set of \emph{fixed} parameters and $\alpha$, the chemoattraction strength, is the free continuation parameter. 
\FixRevOne{The vector $\mathbf{f}$ represents the temporal derivatives, and it is obtained via standard \emph{Galerkin} FEM discretisation.
}

\dsdkFive{Winter et al.}\FixRevFive{\citet{winters1990tracking}} determine the next bifurcation point and its corresponding bifurcation pattern as follows. 
They first initialise the system with \FixRevOne{the only none-zero homogeneous steady-state solution} $\mathbf{x_0}\equiv (a_0 = N,\,b_0 = \rfrac{N}{1+N})$ and $\alpha = \alpha_0$ as a guess for the next bifurcation point. Next, using a \emph{Newton method}, they solve an \emph{extended system} of equations $\mathbf{f}(\mathbf{x}, \mathbf{p}, \alpha) = 0,$ and $\mathbf{J} \Delta \mathbf{x} = 0$ with $\mathbf{x} = \mathbf{x_0} + \Delta \mathbf{x}$ and $\alpha = \alpha_0 + \Delta \alpha$ as the unknowns, and $\mathbf{J}$ as the Jacobian of $\mathbf{f}$ \FixRevThree{at $(\mathbf{x}_0, \mathbf{p}, \alpha_0)$}. \FixRevOne{The equation extension solves for a non-trivial $\Delta \mathbf{x}$  that lies in the nullspace of the Jacobian matrix and thus represents a pattern that may grow without affecting the system steadiness}. From a solution of this extended system, $(u^b,\,v^b) = \Delta \mathbf{x}$ represents the emergent bifurcation pattern and $\alpha^b =  \alpha_0 + \Delta \alpha$ defines the corresponding bifurcation point along the trivial branch. \FixRevOne{Then, they use a \emph{pseudo-arclength} continuation approach to follow the new branch and to determine the family of steady-state solutions along it.} 

Our framework avoids the repetitive complexity of solving an extended system by directly locating each bifurcation point. Also, more importantly, our approach works on arbitrary surfaces and not just on rectangles. Consider linearising Equation~\ref{eqn:murray_nonlin} near the homogeneous solution $a_0 = N,\, b_0 = N/(1+N)$.
\FixRevOne{We find the terms in the generic linearised PDE from Equation~\ref{eqn:gen_cross_rd_uv_lin} as} 
\begin{align}
&{}^uD_u=D,\quad{}^uD_v= -\alpha N,\quad{}^vD_u=0,\quad{}^vD_v=1,\nonumber\\
&{}^uK_u= - C N S,\,\,^uK_v=0,\,\,^vK_u= \rfrac{S}{(1+N)^2}\,\text{ and }\,^vK_v= - S\,.
\end{align}  
We substitute these parameters in Equation~\ref{eqn:lambda_constraints} for a given mode $\phi_i$ with eigenvalue $\lambda_i = \Lambda$ to express the continuation parameter $\alpha$ at the bifurcation point as
\begin{align}
\alpha^b\,\,=\,\,(1 + N)^2 \left[\,\, C\,\left(1 + \frac{S}{\Lambda}\right)\,\,+ \frac{D}{N}\,\left(1 + \frac{\Lambda}{S}\right) \,\,\right]\,. 
\label{eqn:murray_alpha}
\end{align}
Making similar substitutions in Equation~\ref{eqn:bif_constraints_cross_diff} gives us the scale factor $s_\Lambda = v_i/u_i$ as
\begin{align}
s_\Lambda\,\,=\,\,\frac{ D \Lambda \,+\, C N S}{ \alpha^b N \Lambda}\,.
\label{eqn:murray_scale_factor}
\end{align}
In our framework, we directly compute an eigenvector $\mathbf{b}_k$ of the discrete Laplacian using Equation~\ref{eqn:DisreteLapGenEigProblem}, corresponding to a continuous eigenfunction $\phi_k$, and its respective eigenvalue $\Lambda$. Then, using Equations~\ref{eqn:murray_alpha} and~\ref{eqn:murray_scale_factor}, we immediately locate the bifurcation point $\alpha = \alpha^b$ and compute the bifurcation pattern $\mathbf{x}_b =(\mathbf{u}_b = \mathbf{b}_k,\, \mathbf{v}_b = s_\Lambda \mathbf{b}_k)$ by setting $u_k=1$, without loss of generality. 
 \FixRevThree{For all experiments with Murray's model, we set $D=0.25$, $C = 1.522$, $N=1$, and $S=1$.}

\subsubsection*{Emergent Patterns.}
\label{ssec:bif_patterns_resuts}

Again, we first validate our framework with analytically known emergent patterns on a simple geometry. We provide a general description of our experiments here, and provide detailed numerical results in the supplemental material (SM02.A3).

\FixRevFive{For the rectangular domain $\Omega$ that we use in this case study, eigenvalues of the LB operator may have geometric multiplicity (independent eigenmodes) greater than one. In such cases, the uniqueness of numerically computed eigenvectors is not guaranteed. Thus, to analyse the accuracy of a numerically computed emergent pattern $\mathbf{x}_b = (\mathbf{u}_b, \mathbf{v}_b)$, we need to first determine an analytically defined vector $\widehat{\mathbf{u}}_b$ which is nearest to $\mathbf{u}_b$. We do this with an approach similar to that of \citet{reuter2009discrete}, i.e., by projecting}
\dsdkFive{We analyse the accuracy of a numerically computed emergent pattern $\mathbf{x}_b = (\mathbf{u}_b, \mathbf{v}_b)$ using a spectral decomposition approach similar to that of Reuter et al (2009). We project}the normalised vector $\mathbf{u}_b$ onto a space with analytically defined basis vectors $\lbrace\mathbf{b}_i\rbrace$, which represent the ground truth patterns. This yields the closest analytically defined pattern $\widehat{\mathbf{u}}_b$. We then use the root mean square error $\epsilon_{RMS} =  \rfrac{||\mathbf{u}_b -\widehat{\mathbf{u}}_b||}{\sqrt{n}}$, with $n$ FEM nodes, as our error measure for the numerically determined bifurcation pattern. We obtain each basis vector $\mathbf{b}_i$ by discretizing its corresponding continuous eigenfunction $\phi_i$ at the FEM node positions. The analytical eigenfunctions on a rectangular domain are   
\begin{align}
&\phi_{i} \,\,= \,\,\cos{(\rfrac{p\pi x}{W})}\cos{(\rfrac{q\pi y}{H})}\,,\text{ and} \quad \quad
\lambda_{i} \,\,= \,\,\left(\frac{p\pi}{W}\right)^2 + \left(\frac{q\pi}{H}\right)^2 \,,\nonumber\\
&\text{ for } \quad\Omega\,\,= \,\,\lbrace\left(x,y,0\right)\,\,| \,\,0 \leq x \leq W = 1,\, 0 \leq y \leq H=4\rbrace\,,
\label{eqn:murray_eigen_func}
\end{align}      
where $W$ and $H$ are the width and height of the rectangle, respectively. 
The index $i$ denotes a wavemode with $p$ sinusoidal extrema (crust/trough) along the $x$--axis and $q$ extrema along the $y$--axis.

\FixRevOne{In the supplemental material we evaluate RMS errors $\epsilon_{RMS}$ for $100$ emergent patterns with our proposed method and the reference method. We assign each emergent pattern an index based on its nearest analytically determined eigenvector $\mathbf{b}_i$. We solve for the eigenvectors with a convergence limit on the order of $10^{-12}$, and this results in low RMS errors in general (on the order of $10^{-4}$\textendash$10^{-6}$). In general, the emergent patterns $\mathbf{u}_i$ found with our method and their projections $\widehat{\mathbf{u}}_i$ onto the analytical solutions are visually indistinguishable. In addition, while the reference method misses out on some emergent patterns due to multiple bifurcations or failures \dsdk{due to}\FixRevThree{of} the chosen test function, we discover all emergent patterns. We also analyse the relative error in locating a bifurcation point $\alpha^b$ corresponding to $\mathbf{x}_b$ by comparing it with an analytically defined $\alpha_{ref}$. We compute $\alpha_{ref}$ from Equation~\ref{eqn:murray_alpha} using analytically defined $\lambda_i$ corresponding to basis vector $\mathbf{b}_i$ closest to $\mathbf{u}_b$. 

We further investigated the impact of the FEM order on the accuracy of the discovered patterns, and we obtained similar conclusions as for the Brusselator model.
Notably we observe that for FEM of order one, the error in locating a bifurcation point grows with the magnitude of the eigenvalue for the respective pattern. However, using higher order FEMs improves results considerably, albeit at the cost of some increase in relative errors for the emergent patterns. See supplemental material (SM02.A2) for details. It also includes an evaluation of the impact of mesh triangulation on the accuracy of our proposed method. 

}

\begin{table*}[t]
\renewcommand{\arraystretch}{1.0}
\begin{minipage}{0.45\textwidth}
\includegraphics[width=\linewidth]{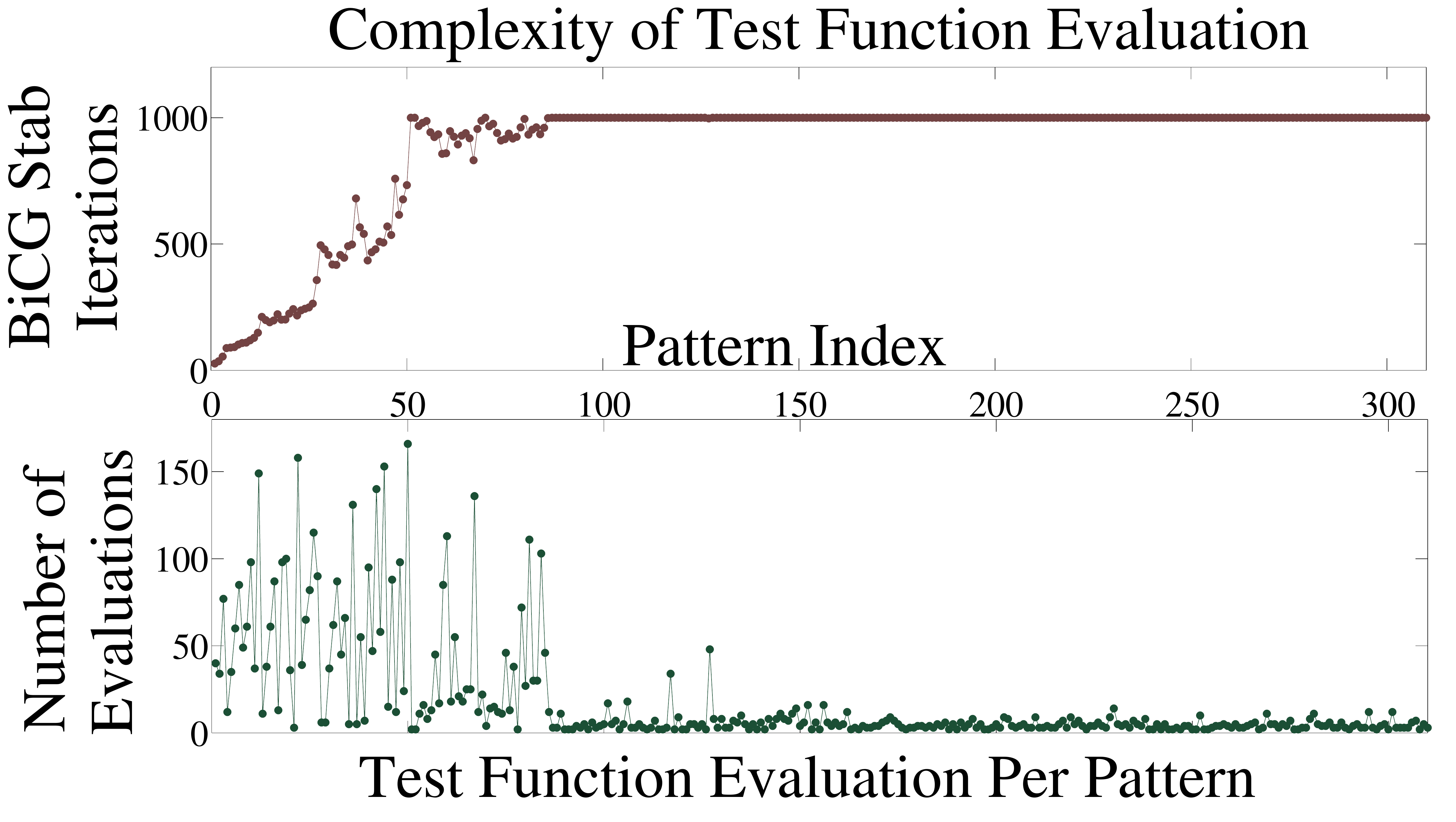}
\end{minipage}\hfill{}
\begin{minipage}{0.5\textwidth}
\begin{tabular}{ m{0.8in}  >{\centering}m{0.4in} >{\centering\arraybackslash}m{0.4in}}
\toprule
\multirow{2}{*}{Computation} & \multicolumn{2}{m{0.8in}}{\centering Time Per Pattern }\\
\cline{2-3} 
 & Reference & Proposed \\
\midrule
Test Function & \multicolumn{1}{c}{ $16.318$s} & \textemdash \\

Bifurcation Point & $3.727$s  & \textemdash \\
3.7271
Emergent Pattern &  \textemdash & $0.750$s \\
\midrule
Total & $20.045$s & $0.750$s \\
\bottomrule
\end{tabular}
\end{minipage}
\caption{Performance measures for Murray's model on a rectangular domain. We used first order FEM with about $4400$ nodes. Graphs on the left show statistics for test function evaluations for the reference method. }
\label{tab:bruss_overheads}
\end{table*}
\FixRevThree{
Table~\ref{tab:bruss_overheads} show a performance comparison between our proposed method and the reference method. We measured average time for different tasks while computing emergent patterns for Murray's model acting on a rectangular domain. We used about $4400$ (first order) FEM nodes for this evaluation with convergence limit set to $10^{-11}$ in each case. 
The graph on the left indicates that the required number of test function evaluations and thus the computational cost for the reference method decreases with the wavemode frequency. This happens because the distance between the wavemodes decreases with the eigenfrequency. However, it also implies that after some point the step size for evaluating the test function along the trivial branch would not be small enough to detection all emergent modes. 
Also note that the complexity of evaluating a test function increases with the wavemode frequency. This complexity is somehow artificially limited to $1000$ iterations since we limit the \emph{BiCG Stab} solver for computing the test function to $1000$ iterations. Overall, Table~\ref{tab:bruss_overheads} indicates that our proposed method is about $25\times$ faster than the reference method in this case.}
 
\begin{figure*}
\centering
        \begin{subfigure}[t]{0.44\linewidth}
	        \centering
                \includegraphics[width=0.8\textwidth]{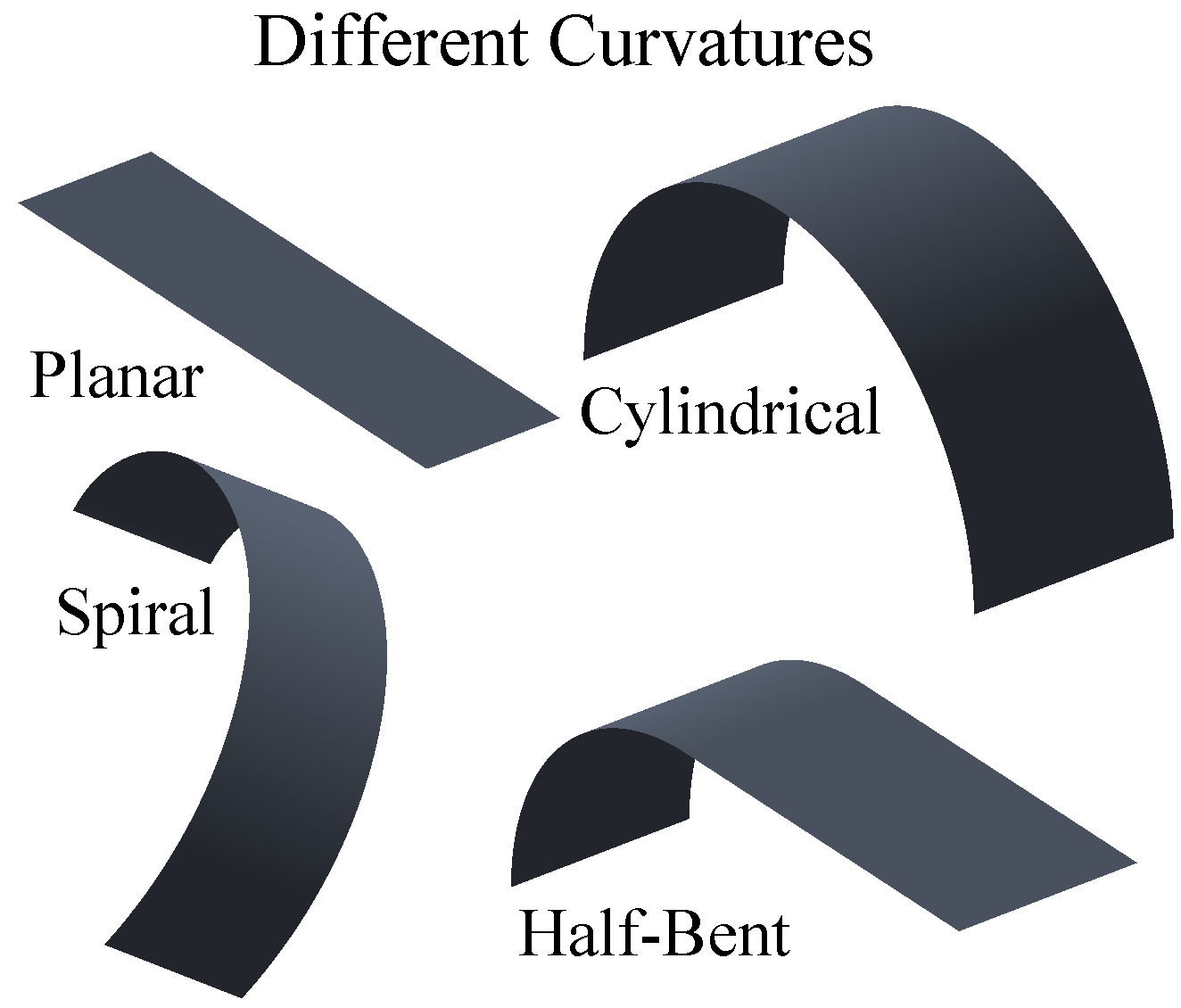}
                \caption{Different developable deformations of the $1 \times 4$ rectangular domain $\Omega$ of equal size (Illustrations are not on the same scale).}
                \label{fig:murray_pattern_errors_diff_curvature_structures}
        \end{subfigure}\hfill{}
        \begin{subfigure}[t]{0.55\linewidth}
					\includegraphics[width=\linewidth]{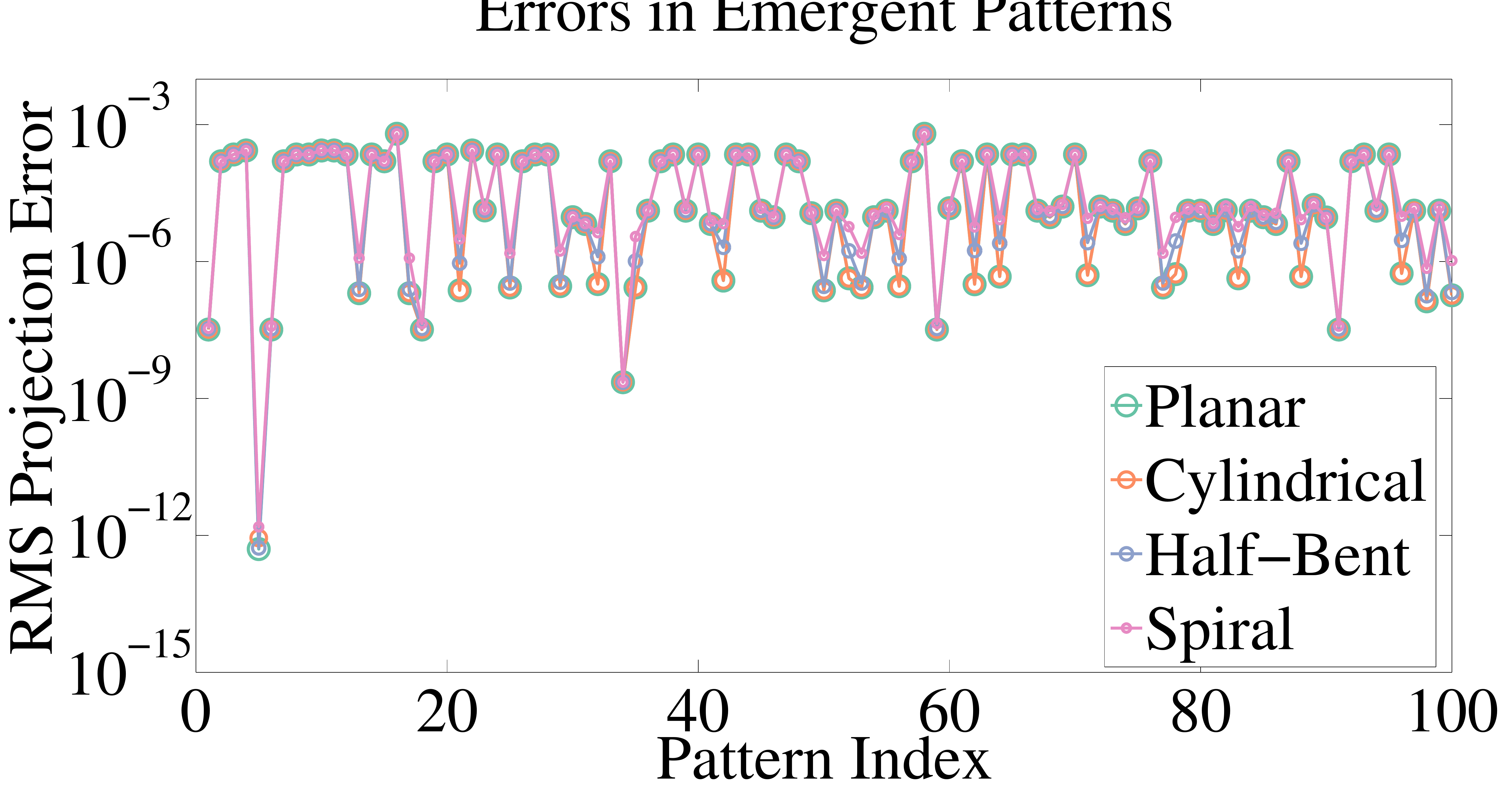}
                \caption{Error plots on a semi-log scale.}
                \label{fig:murray_pattern_errors_diff_curvature_errors}
        \end{subfigure}%
\caption{RMS projection error in computing the wave mode pattern $\phi_i$ using our method on developable surfaces with different curvature profiles for Murray's model.}
\label{fig:murray_pattern_errors_diff_curvature}
\end{figure*}
Next, we evaluate the impact of different domain shapes on emergent patterns found with our framework. Figure~\ref{fig:murray_pattern_errors_diff_curvature_structures} shows three different \emph{developable} deformations of a $1 \times 4$ rectangular domain $\Omega$. This includes a \emph{cylindrical} surface with constant curvature, a \emph{half-bent} surface composed of two parts with constant curvatures and curvature discontinuity between them, and a \emph{spiral} surface with a continuously varying curvature profile. For comparison, we also used the original $1\times4$ rectangular domain $\Omega$. 
We used a regular grid of about $4400$ vertices in each case. With developable deformations, we do not expect that the non-planarity of the domains has an influence on emergent patterns. Figure~\ref{fig:murray_pattern_errors_diff_curvature_errors} shows error statistics for all three deformed surfaces along with the rectangular surface as a reference. The differences in errors for different curvature profiles appear to be well within general variations in error statistics. We found a similar trend in relative errors for computing the bifurcation points and conclude that, as expected, the influence of developable deformations is negligible.

\begin{figure}
\centering
        \begin{subfigure}[t]{\linewidth}
                \includegraphics[width=\textwidth]{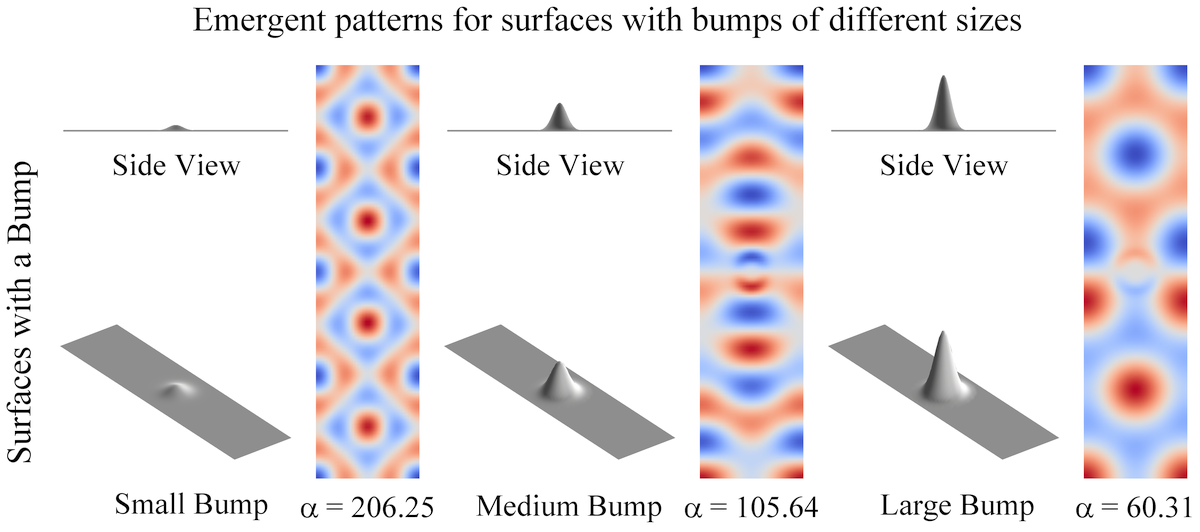}
                \caption{Bump at the center. \dsdk{\color{NoteC} WE CAN REMOVE THIS FIGURE, if required}}
                \label{fig:murray_patterns_on_center_bump}
        \end{subfigure}\\
        \begin{subfigure}[t]{\linewidth}
\begin{tikzpicture}
\node[anchor=south west,inner sep=0] (image) at (0,0) {
                \includegraphics[width=\textwidth]{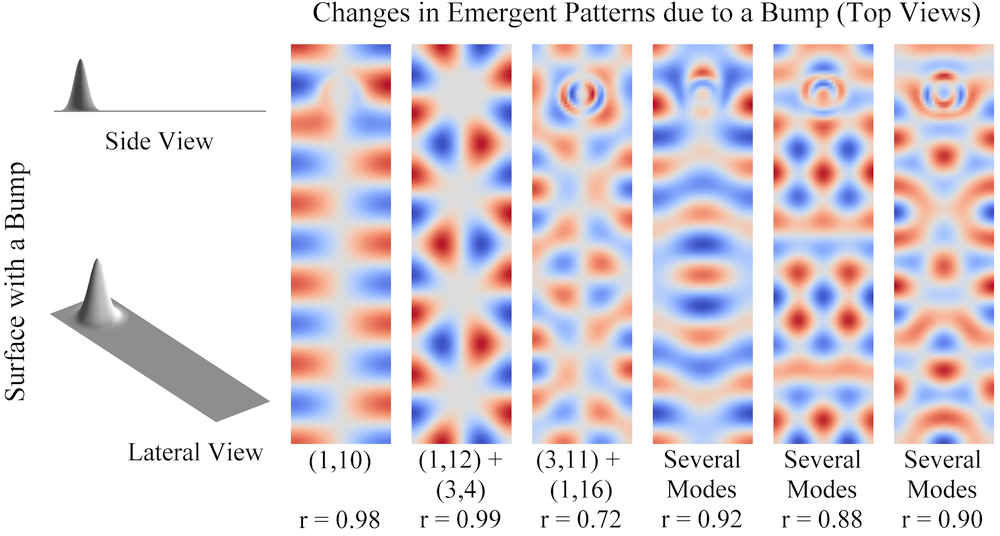}
};
\begin{scope}[x={(image.south east)},y={(image.north west)}]
    \end{scope}
\end{tikzpicture}            
                \caption{Bump at the edge.}
                \label{fig:murray_patterns_on_edge_bump}
        \end{subfigure}%
\caption{Changes in emergent patterns for Murray's model due to a surface bump.}
\label{fig:murray_patterns_with_bumps}
\end{figure}

To study the influence of local non-developable surface deformations on emergent patterns, we use surfaces with bump deformations at different locations or size to the reference rectangular domain $\Omega$, see Figure~\ref{fig:murray_patterns_with_bumps}. 
These bumps are akin to the emergence of animal limbs or appendices. We observed that even small local non-developable deformations bring out well-defined global deviations in emergent patterns as shown in the figure. 
\FixRevOne{To obtain an intuition about these new emergent patterns, we map them back onto the flat rectangular domain and project the mapped patterns onto the eigenvectors of the flat domain, as illustrated in Figure~\ref{fig:murray_patterns_on_edge_bump}. A main observation is that we can accurately represent \FixRevThree{many of} the new patterns emerging on the deformed shape with only a few eigenvectors of the flat domain, where we measure accuracy as the correlation between a new pattern and its representation using a few eigenvectors.
We indicate the few dominant wavemodes of the flat domain along with the correlation factor $r$ below the emergent patterns in Figure~\ref{fig:murray_patterns_on_edge_bump}. This leads to the intuition that surface deformation leads to changes of the eigenfunctions and the eigenspectrum, such that \FixRevThree{many} new, complex patterns emerging on deformed surfaces can be understood as linear combinations of simpler patterns on an original, undeformed domain.}

\begin{figure}[t]
\centering
\includegraphics[width=0.9\linewidth]{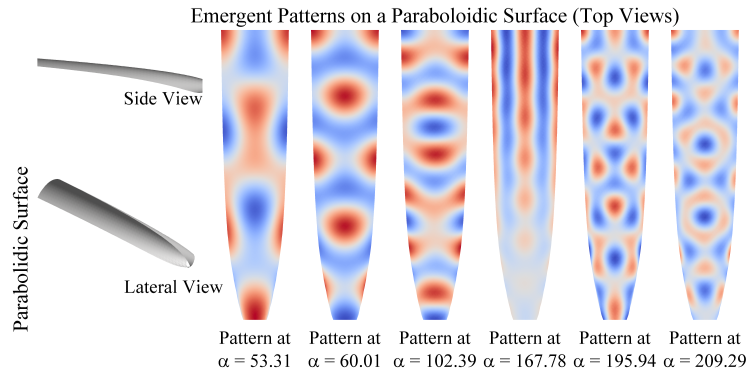}
\caption{Emergent patterns on a paraboloid surface.}
\label{fig:murray_pattern_on_paraboloid}
\end{figure} 
We also examined the influence of global non-developable deformations on emergent patterns. Figure~\ref{fig:murray_pattern_on_paraboloid} shows one such distortion comprised of a \emph{sliced} paraboloid with a perimeter equal to that of the reference rectangular domain. Such a surface is representative of the region on a snake body which is undergoing \emph{morphogenesis} for skin pattern formation. 
Again, we see interesting changes in emergent patterns which appear as deformations and mixing of eigenfunctions for the reference domain. {Thus, we conclude that non-developability in surface geometry plays an important role on pattern formation and must be taken into consideration while studying such emergent patterns}.      

Finally, we applied Murray's model to a $3D$ reconstruction of a real \emph{gecko} lizard body. In this hypothetical case, the eyes, the paws and the ventral side of the body are pruned since these regions do not participate in pattern formation. We used\mzk{a \emph{random uniform}} first-order FEM\mzk{mesh} with about $4600$ nodes\mzk{ to study emergent patterns}. Figure~\ref{fig:murray_pattern_on_lizard} illustrates a subjective comparison of several patterns observed on juvenile geckos and our simulations. For visualization, we perform a nonlinear soft-thresholding on the emergent patterns.     
Our simulation results give a good impression of the variety of patterns observed in nature and make a good case for using actual surface geometry in studying pattern formation.    
\begin{figure*}
\centering
\begin{subfigure}[t]{0.98\textwidth}
\includegraphics[width=0.98\linewidth]{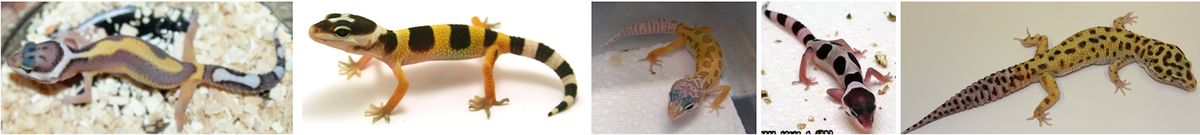}
\caption{Patterns on real lizards.}
\end{subfigure}
\begin{subfigure}[t]{0.98\textwidth}
\includegraphics[width=0.98\linewidth]{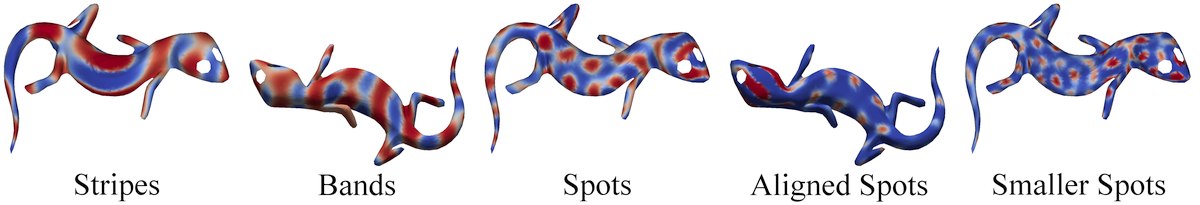}
\caption{Simulated patterns.}
\end{subfigure}
\caption{Emergent patterns on a \emph{Gecko} lizard body surface. }
\label{fig:murray_pattern_on_lizard}
\end{figure*}
\subsubsection*{Branch Tracing}
\label{ssec:branch_tracing_resuts}
{\color{AddedNew4}

\begin{figure}[t]

\centering
\begin{tikzpicture}
\node[anchor=south west,inner sep=0] (image) at (0,0) {
\includegraphics[width=0.9\linewidth]{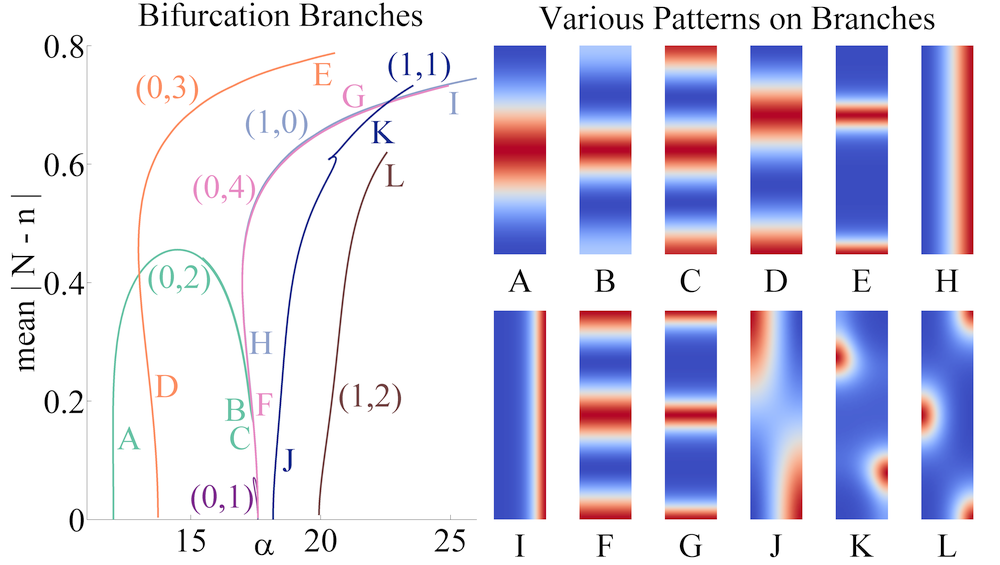}
};
\begin{scope}[x={(image.south east)},y={(image.north west)}]
    \end{scope}
\end{tikzpicture} 
\caption{Branch tracing results for rectangular domain $\Omega$.}
\label{fig:murray_branches_and_patterns}

\end{figure}
We present results for branch tracing as explained in Section~\ref{sssec:BranchTracing} with Murray's chemotactic model. 
We use first-order FEM in all experiments with $\alpha$ as the continuation parameter and other parameters fixed \FixRevOne{as before}. In general, for computing the eigenvectors we use a convergence limit $\epsilon \in [ 10^{-8},\,\, 10^{-11}]$ on the root-mean-square residual error for solving Equation~\ref{eqn:DisreteLapGenEigProblem}. Also, we set the tolerance limit to $10^{-12}$ for the AztecOO inner loop GMRES solver. We perform branch switching with the parallel programming distance  $\epsilon_0 \in [ 1e{-3}, 0.07 ]$ for Equation~\ref{SM02-eqn:bordering_algo} in the supplemental material (SM02). Also, the divergence limit $\alpha_\epsilon$ discussed in Section~\hspace{1.2cm}\ref{SM02-sssec:BranchTracing} in the supplemental material (SM02) is set in the range $[ 10^{-4}, 10^{-2}]\times \alpha_0$. For an arclength continuation step, we set the convergence limit in the range $[ 1e{-8},\,\, 1e{-11}]$ for the RMS residual error.    

\begin{figure}[t]
\centering
\begin{tikzpicture}
\node[anchor=south west,inner sep=0] (image) at (0,0) {
\includegraphics[width=0.9\linewidth]{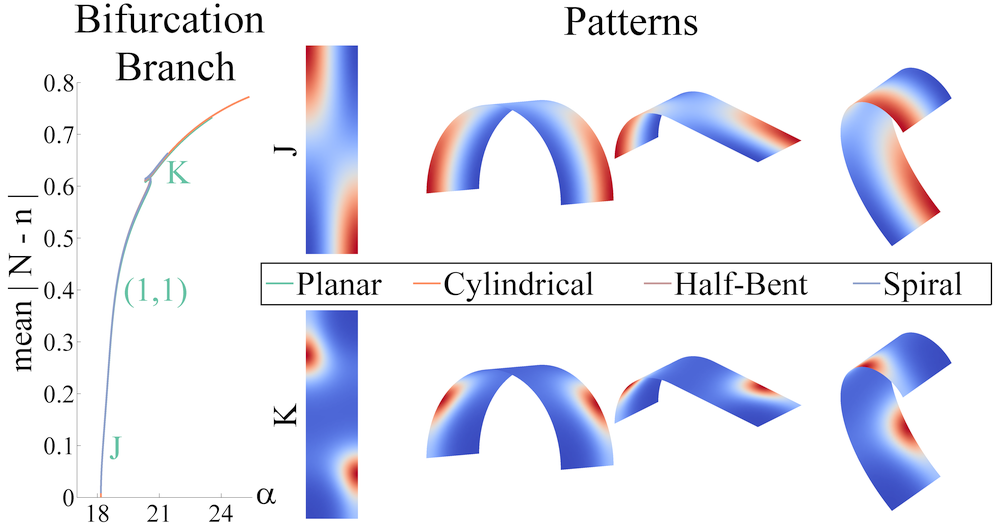}
};
\begin{scope}[x={(image.south east)},y={(image.north west)}]
    \end{scope}
\end{tikzpicture} 
\caption{Branch $J$\textendash$K$ for various developably deformed domains and corresponding patterns in sections $J$ and $K$ for the branches.}
\label{fig:murray_branch_patterns_and_curvature}
\end{figure}
Figure~\ref{fig:murray_branches_and_patterns} shows the branches traced for a rectangular domain $\Omega$ using our proposed direct approach. We first compute the eigenvectors for the domain with about $4400$ FEM nodes. We then directly compose bifurcation patterns, compute the respective bifurcation points, and perform branch tracing as explained in Section~\ref{sec:framework}. Our results for branches and segments $A$ \textendash $K$ in Figure~\ref{fig:murray_branches_and_patterns} are qualitatively similar to those presented by \citet{winters1990tracking}. Unlike \citet{murray1991pigmentation} who change the underlying geometry for the rectangular domain to separate out multiple branches $H$\textendash$I$ and $F$\textendash$G$, we trace these branches using our cosine factor method for branch switching \FixRevFour{(Refer the second approach in Section~\ref{sssec:BranchTracing})} without changing $\Omega$. Our framework thus enables experiments with different linear combinations of multiple bifurcation modes as seen later. Next, we repeat the branch tracing experiment for developable surfaces with different curvatures to obtain qualitatively similar results. For comparison, we present branch $J$\textendash$K$ for all four domains and resolved patterns in sections $J$ and $K$ of the branch, in Figure~\ref{fig:murray_branch_patterns_and_curvature}. 

\begin{figure*}
\color{IntroC}
\centering
\includegraphics[width=0.98\linewidth]{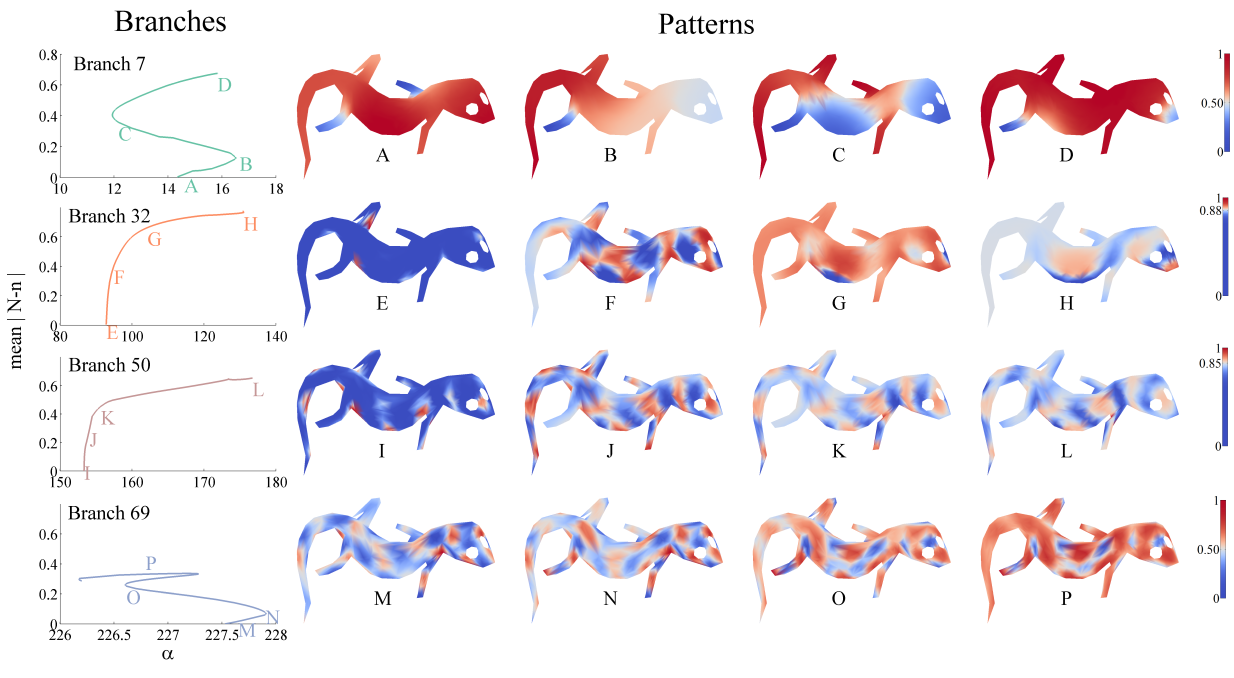}
\caption{Tracing branches on an arbitrary surface representing a \emph{Gecko} lizard. Each row presents progressive changes in the emergent pattern as we move along its respective branch.}
\label{fig:murray_gecko_branches}
\end{figure*}
After validating our framework with known results for simple geometries, we trace branches for an arbitrary surface geometry representing the \emph{Gecko} lizard surface with about 900 FEM nodes. 
We trace branches for the first seventy eigenvectors of this shape. Figure~\ref{fig:murray_gecko_branches} shows some of the interesting branches across the explored spectra. The leftmost column in Figure~\ref{fig:murray_gecko_branches} plots the branches and each row shows the evolution of a pattern along its respective branch. We perform a nonlinear soft-thresholding again to visualise each pattern.
 \FixRevOne{For qualitative comparison, we use the same nonlinear color mapping for all the solutions along a branch after normalizing the pattern range to $\left[0,1\right]$ (as shown in the rightmost column of the figure).} 
The first row in the figure shows one of the \emph{low frequency} branches (branch $7$). We notice clear qualitative changes that occur progressively as we move 
along the branch. 
The same is true for other branches such as two \emph{mid-frequency} branches (branch $32$ and $50$) and one \emph{high-frequency} branch (branch $69$) shown here\footnote{We label the branches as low, mid or high frequency branches based on the range of the eigen-spectra which is explored.}. Many interesting patterns such as $C$, $G$, $H$, $L$ and $P$, which are qualitatively quite different from the emergent patterns, are discovered with branch tracing. Thus, we can say that branch tracing contributes significantly to the exploration of the solution space of a nonlinear system of PDEs and it is an important tool to study such biological systems. 
\FixRevOne{Unlike Murray's approach for branch tracing, our framework has an added advantage of tracing branches for composite emergent patterns. 
In Figure~\ref{fig:murray_multiple_bifurcation} we show results for a composite emergent pattern on the rectangular domain $\Omega$. 
This domain has three emergent patterns at $\alpha = 17.575$. Two of those patterns are shown as $H$ and
\begin{figure}
\centering
\includegraphics[width=0.8\linewidth]{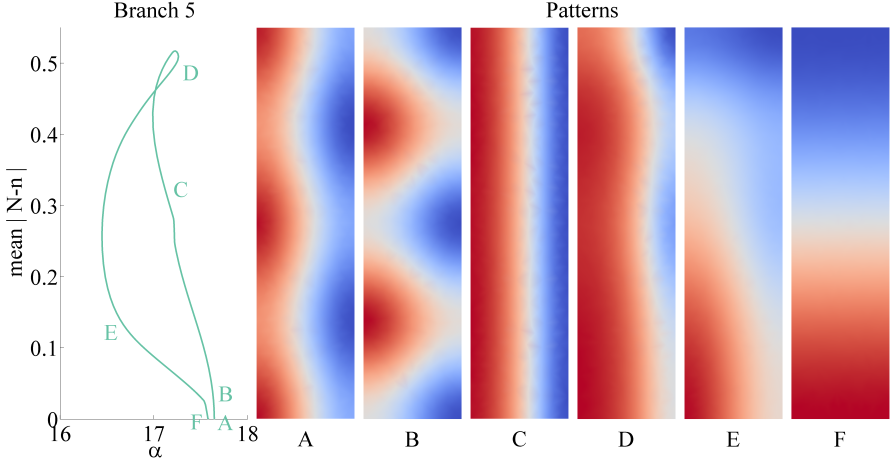}
\caption{Tracing branch for a composite pattern.}
\label{fig:murray_multiple_bifurcation}
\end{figure}
 $F$ in Figure~\ref{fig:murray_branches_and_patterns}. We take a linear combination of patterns in $H$ and $F$ and trace a branch for it, as shown in Figure~\ref{fig:murray_multiple_bifurcation}. Pattern $A$ in the figure shows the starting pattern. A variety of patterns appear as we move along the branch, changing progressively from $A$ to $F$. All the observed patterns appear to be mainly composed of three emergent patterns at the multiple bifurcation point. Thus, this branch illustrates that such linear combinations could result in steady patterns in nonlinear regions far away from the trivial branch.   

}
\subsubsection*{Results for Multiresolution Adaptation}
\label{ssec:multireso_results}
\color{AddedNew4}

\begin{figure}
\centering
\begin{subfigure}[b]{0.5\linewidth}
\centering
                \includegraphics[height=0.6\textwidth]{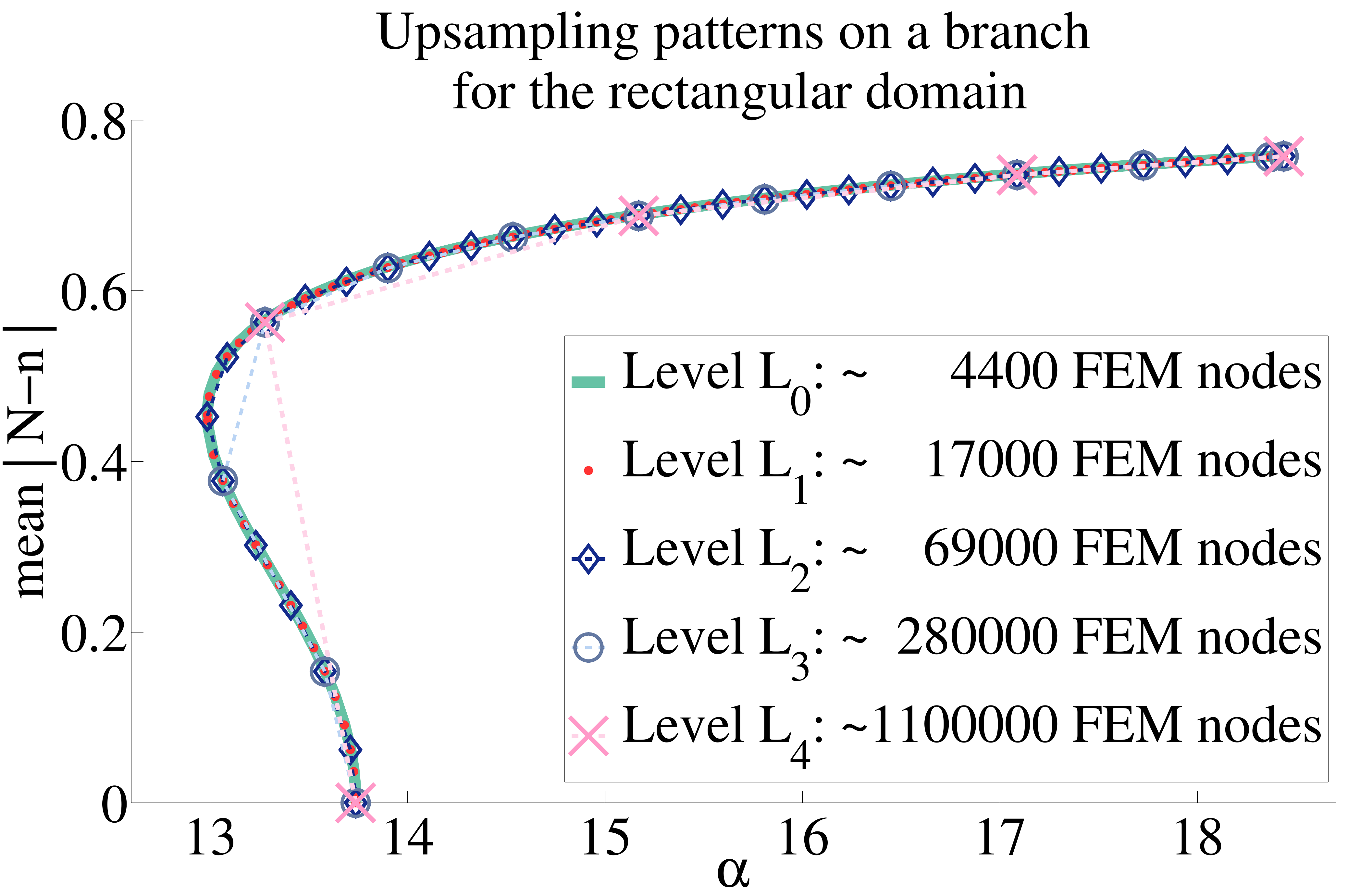}
                \caption{Branch}
                \label{fig:upsampling_branch_for_rectangle}
        \end{subfigure}%
\begin{subfigure}[b]{0.5\linewidth}
\centering
                \includegraphics[height=0.6\textwidth]{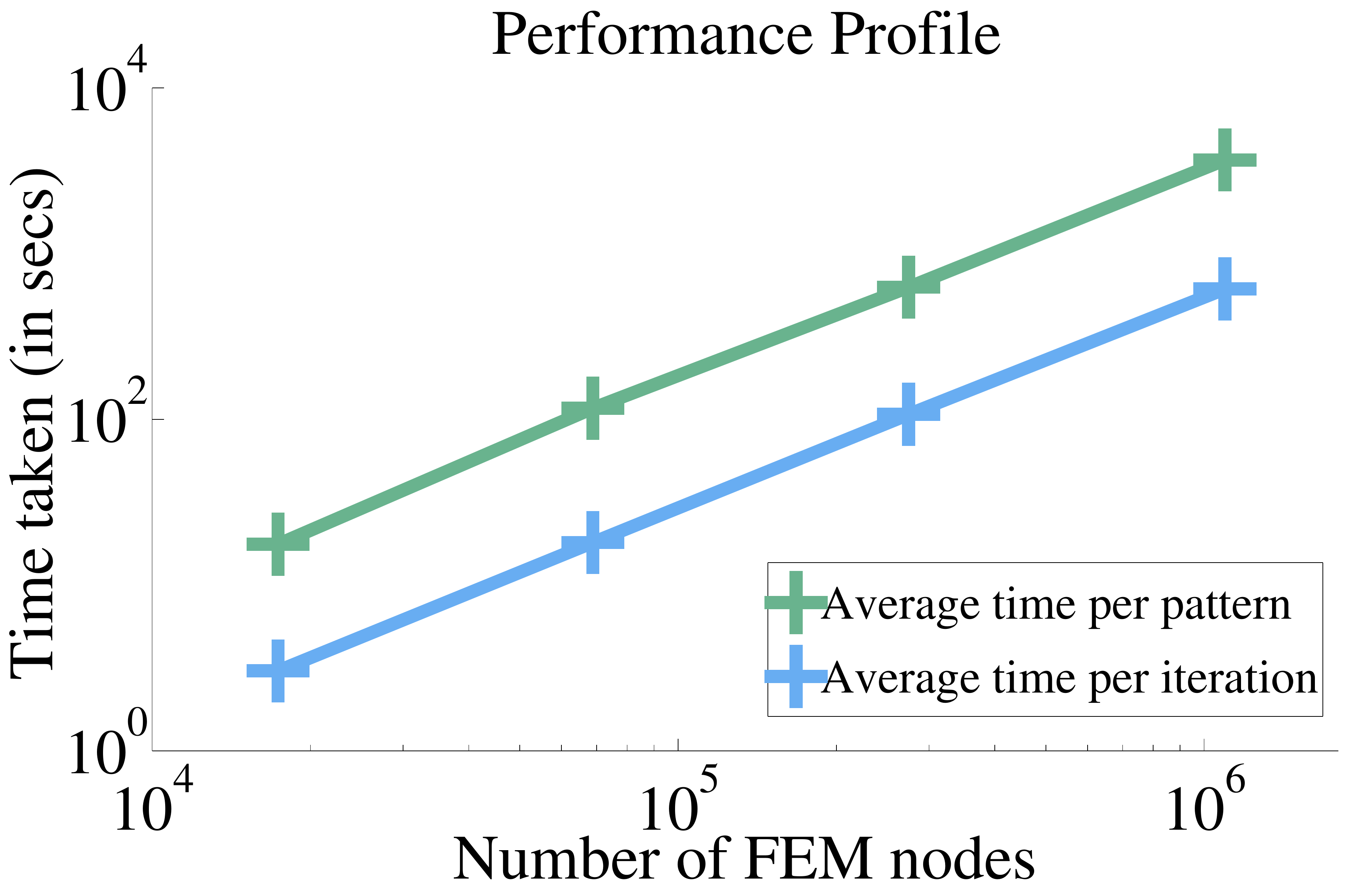}
                \caption{
                Performance plot on a \emph{log-log} scale}
                \label{fig:upsampling_performance}
        \end{subfigure}%

\caption{An upsampled branch for the rectangular domain $\Omega$ and plots for performance analysis of this upsampling.}
\label{fig:upsampling_branches}
\end{figure}
Our progressive geometric multigrid approach (Section~\ref{sec:adaptation}) allows us to perform branch tracing at high resolutions of up to one million FEM nodes. We proceed in two steps, namely (a) resolution improvements and (b) geometric improvements through multiple levels. 
We first demonstrate the upsampling of patterns on a branch for the rectangular domain in Figure~\ref{fig:upsampling_branches}. Since this domain is planar, we only need to perform step (a) for our multiresolution approach in this case. Figure~\ref{fig:upsampling_branch_for_rectangle} shows the upsampling of a branch originating at $\alpha = 13.736$ (branch $D$-$E$ in Figure~\ref{fig:murray_branches_and_patterns}) from level $L_0 \approx 4400$ FEM nodes to level $L_4 \approx 1.1$M FEM nodes. From one lower level to the next, we divide each quad finite element into four quad elements. For first order FEM this increases the number of FEM nodes by about $4\times$. Thus, with four levels of higher resolution, we go from about $4400$ to $1.1$ million FEM nodes.

The flexibility of upsampling only a selective subset of solutions can help to reduce the computational load of such studies, and it allows the user to prioritise the upsampling of selected solutions. In Figure~\ref{fig:upsampling_branch_for_rectangle} we upsampled only about one third of the solution patterns from each level to the next. The figure shows that the solution patterns at each level lie close to the original branch at the lowest level. We found that the upsampled patterns resolved well in a few iterations and appeared subjectively very similar across all levels (not shown here for brevity). 
We also profiled the performance of our framework for this example. Table~\ref{tab:upsampling_on_rect} shows the average number of iterations required to achieve convergence at each level along with the time taken. Figure~\ref{fig:upsampling_performance} shows a \emph{log-log} plot of average time taken for each iteration and total time for convergence of a pattern over the number of FEM nodes. Both plots are near linear in nature as illustrated by the line fit. Both fitted lines have a slope of $1.281$, indicating the computational complexity of upsampling a given solution as $\mathcal{O}(N^{1.281})$ with $N$ FEM nodes. \FixRevThree{In contrast to our method, we found that a multi-level algebraic grid based preconditioner from the Trilinos library to directly trace a branch for the rectangular domain with about $100$K FEM nodes failed to converge.} 

\FixRevFive{Speaking of parallelisability, for the above case, we estimate that the overhead of the repeated task of reading mesh files and populating the internal data structures is less than $13\%$ of the time taken to resolve a pattern at the highest level ($L_4$). Thus the theoretical limit on the utility of a parallel processor is set high at $\rfrac{1}{1.13} \times 100 = 88.5\%$.}

\begin{table}[htbp]
  \centering
  \caption{Performance profile for upsampling patterns on a branch for Murray's model acting on the rectangular domain $\Omega$.}
  \resizebox{\linewidth}{!}{ 
    \begin{tabular}{rrrrr}
    \toprule
    \multicolumn{1}{c}{\multirow{2}[4]{*}{\textbf{Quantity}}} & \multicolumn{4}{c}{\textbf{Level}} \\
   
    \multicolumn{1}{c}{} & \multicolumn{1}{c}{\textbf{$L_1$}} & \multicolumn{1}{c}{\textbf{$L_2$}} & \multicolumn{1}{c}{\textbf{$L_3$}} & \multicolumn{1}{c}{\textbf{$L_4$}} \\
     \midrule
            &         &         &         &  \\
    \textbf{FEM nodes} & 17359   & 68869   & 274345  & 1095121 \\
    \textbf{} &         &         &         &  \\
    \textbf{Iterations per pattern for convergence (average)} & 5.806   & 6.457   & 5.857   & 6.000 \\
    \textbf{Average time for convergence (in sec)} & 17.600  & 116.582 & 626.523 & 3668.539 \\
    \textbf{Average time per iteration (in sec)} & 3.031   & 18.055  & 106.967 & 611.423 \\
            &         &         &         &  \\
    \bottomrule
    \end{tabular}%
    }
  \label{tab:upsampling_on_rect}%
\end{table}%

\begin{figure*}
  \centering
\begin{tikzpicture}
\node[anchor=south west,inner sep=0] (image) at (0,0) {
   \includegraphics[width=\textwidth]{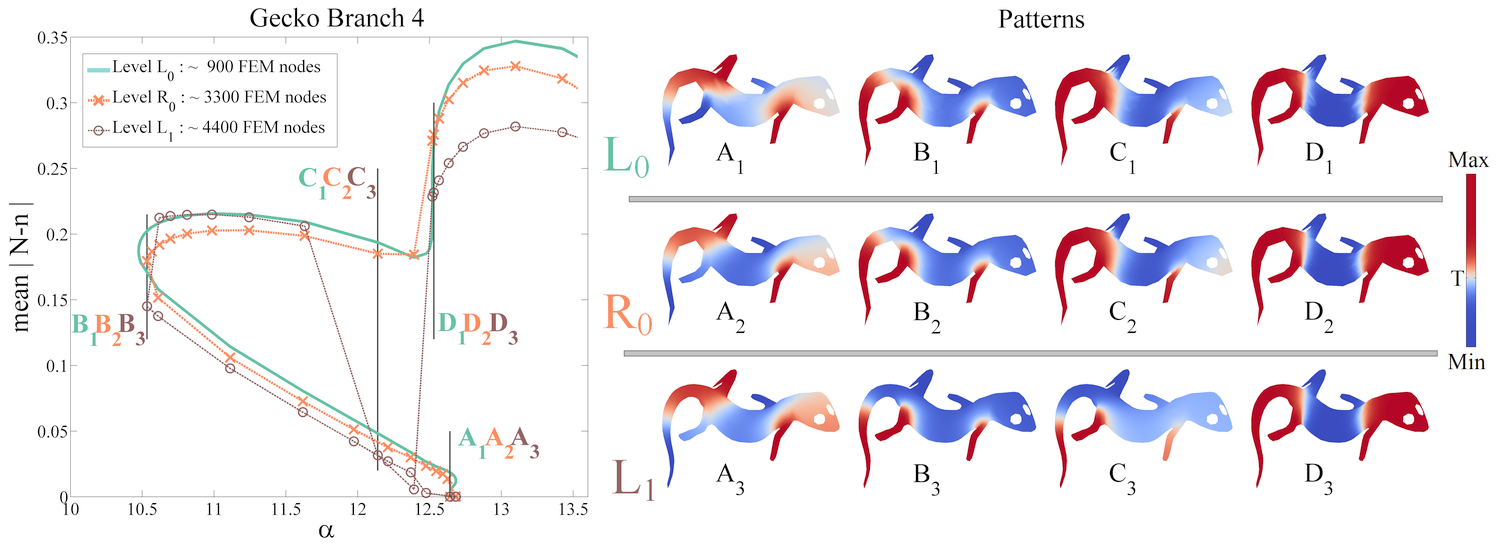}
};
\begin{scope}[x={(image.south east)},y={(image.north west)}]
    \end{scope}
\end{tikzpicture}    
  \caption{Upsampling a branch with low frequency patterns on an arbitrary surface, $\Omega$. We see results for both, resolution improvement from $L_0$ to $R_0$ and geometry improvement $R_0$ to $L_1$ in this example. }
  \label{fig:upsampling_branch_4_for_gecko}
\end{figure*}

We next demonstrate upsampling of a branch on the Gecko surface. In this case, we need to perform both (a) resolution improvement as well as (b) geometric improvement in our multiresolution approach. Figure~\ref{fig:upsampling_branch_4_for_gecko} illustrates these two steps for upsampling patterns up by one level. We trace a branch at the lowest level  $L_{0}$ (green-coloured branch in the figure) and show qualitatively different patterns along the branch as $A_1$ - $D_1$\footnote{In this example as well, we apply nonlinear soft-thresholding to aid visualization of the patterns. We use the same nonlinear mapping for each pattern which is first scaled and off-setted to the range $[0,1]$.}. 
We first upsample the patterns to level $R_0$, which has a higher resolution but the same geometry as $L_0$. 
We plot the branch  at level $R_{0}$ (orange branch in the figure) and notice that it drifts slightly from the original branch. However, we find patterns  $A_2$ - $D_2$ to be qualitatively similar to the corresponding patterns on the branch for level $L_0$. Finally, we upsample the patterns to the improved geometry at level $L_1$. Here again, we find considerable changes in the plot for the branch while the patterns do not change much qualitatively, except for $C_3$ (see patterns $A_3$ \textendash  $D_3$.). The qualitative changes are limited across the levels mostly because of the low-frequency nature of the patterns. At the same time, changes in the branch contours indicate that the resolution and geometry of the surface domain can influence the results. Particularly, $C_3$ changes considerably because there is a branch-jump from $C_1$ to $D_1$ at the lowest level which gets corrected with our upsampling. We discuss branch-jumping and other issues in detail in Section~\ref{ssec:challenges_limitations}. 

\begin{figure*}
\color{IntroC}
\centering
\begin{tikzpicture}
\node[anchor=south west,inner sep=0] (image) at (0,0) {
\includegraphics[width=\linewidth]{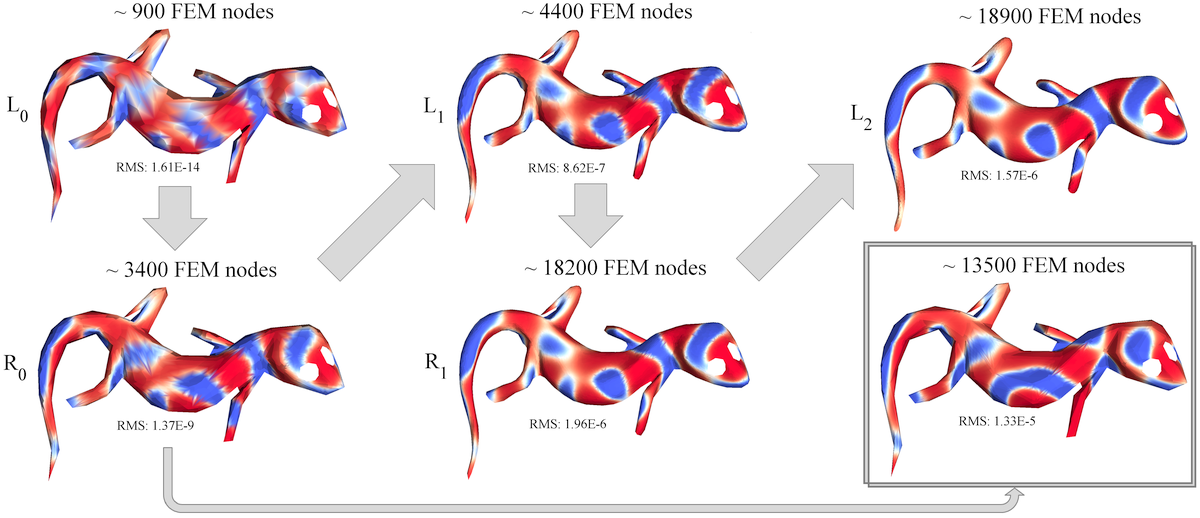}
};
\begin{scope}[x={(image.south east)},y={(image.north west)}]
    \end{scope}
\end{tikzpicture}    
\caption{Upsampling a pattern on a mid frequency branch for the \emph{Gecko} surface in a far-off nonlinear region for the solution space of the underlying PDEs. We upsample results over two levels until patterns are fixed qualitatively. Note the pattern in the bottom right corner with no geometric improvements is qualitatively different from the pattern in the top right with geometric improvements while upsampling.}
\label{fig:upsampling_gecko_branch50}
\end{figure*}
Next, we present an example of upsampling a pattern on a \emph{mid frequency} branch (branch $50$ in Figure~\ref{fig:murray_gecko_branches}), up by two levels. Figure~\ref{fig:upsampling_gecko_branch50} shows the results for levels $L_0$ - $L_2$. For qualitative comparisons across different geometries, we map the values of each pattern into the range $[0,1]$. 
Level $L_0$ has about $900$ FEM nodes, and the pattern looks unresolved even with a low RMS error of $1.61\times 10^{-14}$ units. We first improve its resolution to level $R_0$ with about $3400$ nodes. Adding more nodes changes the pattern qualitatively in several regions \FixRevThree{like the tail and the right forelimb}. 
Next, we improve the geometry and resolve the pattern at level $L_1$ with about $4400$ nodes. Again we observe qualitative differences from level $R_0$ in several regions \FixRevThree{like the right limbs}. 
We further upsample the pattern to levels $R_1$ and finally $L_2$, and find that the pattern at level $L_2$ with about $18900$ FEM nodes is qualitatively very similar to that at level $L_1$. We thus conclude that the pattern is well resolved, we do not need to further upsample it and the geometry at level $L_1$ sufficed to resolve the pattern. 

For comparison, we also continue to sub-divide the quad elements from level $R_0$ for one more step, while keeping the geometry from $L_0$, to resolve the pattern with about $13500$ nodes. \FixRevOne{We show the result in the bottom-right corner, and observe that the resolved pattern is qualitatively different from those for the improved geometries $L_1$ and $L_2$. This illustrates the effect that small changes in geometry can have on the resulting patterns.

\begin{figure*}
\color{IntroC}
\centering
\begin{tikzpicture}
\node[anchor=south west,inner sep=0] (image) at (0,0) {
\includegraphics[width=\linewidth]{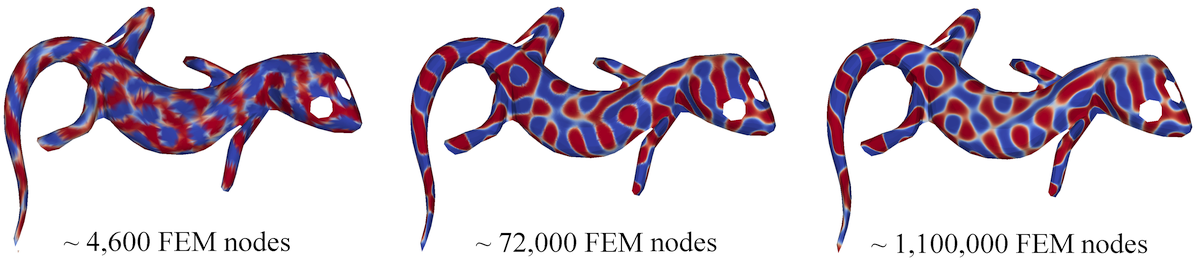}
};
\begin{scope}[x={(image.south east)},y={(image.north west)}]
    \end{scope}
\end{tikzpicture}    
\caption{Upsampling an emergent pattern on a \emph{Gecko} surface for a resolution up to $1M$ FEM nodes. We begin with a relatively good surface geometry \FixRevOne{of about $4600$ FEM nodes}, which we do not refine during upsampling in this example.}
\label{fig:upsampling_gecko_high_freq}
\end{figure*}

Finally, Figure~\ref{fig:upsampling_gecko_high_freq} shows our upsampling results for a very high frequency emergent pattern (eigenvector $250$). Starting from a geometry with about $4600$ nodes, we upsample by only increasing the mesh resolution, but without changing the geometry. We show the results after two and four steps of upsampling in the middle and on the right, respectively. All the patterns were resolved with an RMS residual error $\mathcal{O}(10^{-8})$. We observe that as the resolution increases,  the pattern \dsdk{boundaries}\FixRevThree{contours} between the red and blue color regions become \dsdk{well defined}\FixRevThree{more discernible}. 
}

\subsection{Challenges, Limitations, and Future work}
\label{ssec:challenges_limitations}

Our framework is capable of determining all bifurcations along the trivial branch and works well to study emergent patterns and trace branches in general. However, there are certain aspects with room for improvement. Some of these are common challenges for all branch tracing algorithms. For example, it is common for a branch tracing algorithm to inadvertently switch over to another nearby branch in the solution space. ~\citet{seydel2010practical} discusses this problem in detail in Section $4.9$ of his book. The possibility of such \emph{branch jumping} in our framework is reduced to a good measure with the use of a \emph{Tangent Scale Factor} to manoeuvre the direction of the \emph{continuation curve} (refer to Section~\ref{sssec:BranchTracing} for details). However, branch jumping cannot be completely avoided by our framework. A similar issue relating to interference between nearby branches is observed. Here we find that a \emph{sibling} branch for a multiple bifurcation may interfere with continuation of the current branch and produce a pattern with a mix across two branches. Such \emph{crosstalk effects} can be reduced by using a tangent predictor in our framework for branch tracing but not ruled out. Another limitation for our framework is related to the \emph{sampling theorem} and FEM discretisation. While, in theory, we can solve for $N$ eigenvectors for a domain $\Omega$ with $N$ FEM nodes, not all of them are usable. For a simple case of a square planar domain, the sampling theorem dictates that there must be at least twice as many nodes along a dimension as the number of cycles for a \emph{wavemode} along that dimension to model the continuous wave accurately. \citet{jiang2010improvement} discuss this specific limitation in detail in context of their problem relating to \emph {inertia moment analysis}. Thus, while our framework is able to solve for a larger number of eigenvectors, only a fraction of it can be used for branch tracing. The simplest way out is to solve the eigenproblem by increasing the resolution at the lowest level $\Omega_0$. However, this increase cannot be arbitrarily high since we can trace branches at the lowest level $\Omega_0$ only with domain-size $\mathcal{O}(10^4)$ or lower. Another limitation of our framework is that while tracing a branch at a higher resolution with the use of our progressive geometric multigrid approach, we only solve for a solution at a higher resolution for the continuation parameter value given by the respective solution at the lowest resolution. Thus for cases where the branch contour may change with upsampling, the parts of a (higher resolution) branch that lie beyond the continuation parameter range given by the lowest level branch become untraceable. This problem could possibly be addressed by either increasing the order of finite elements to reduce the change in the branch contour with upsampling or by allowing/forcing the continuation parameter to drift while upsampling the solutions at the twists and turns of a branch~\FixRevThree{\citep{bolstad1986multigrid}}. Finally, the convergence of an upsampled solution is limited by the quality of the quad-mesh triangulation for FEM discretisation. \dsdk{This happens because the GMRES inner loop to solve for  a direction vector for a nonlinear solver to upsample the solution works in a least-square sense and this leads to a minima which is not converged to desired accuracy for the residual error.}  
Future work on our framework would focus on some of the challenges and limitations discussed here.

\section{Conclusions}
\label{sec:summary}

We have presented a framework to perform bifurcation analysis for reaction diffusion systems  on arbitrary surfaces. Our framework uses a compositional approach instead of traditional detection approach to discover new emergent patterns along the trivial branch with a homogeneous pattern. We discussed the boundary conditions which make the Laplacian-Beltrami operator acting on a surface domain \emph{Hermitian}. Such an operator is \emph{self-adjoint} and its eigenfunctions form an orthonormal basis set for all surface functions that satisfy given boundary conditions. We derive formulae to directly compose bifurcation patterns for two-component RD systems and to compute their respective bifurcation points. Our derivations substitute a spectral decomposition of the solution into a generalised system of PDEs linearised near homogeneity. Such a generalised system is representative of all two-component RD systems with or without cross-diffusion. Our framework first computes (FEM) discretised eigenvectors for the Laplacian-Beltrami operator acting on a given surface. It then composes bifurcation patterns and computes their respective bifurcation points using our derivations. In addition, our framework supports a multiresolution branch tracing algorithm. We propose a \emph{progressive geometric multigrid} based approach with multiple levels for branch tracing.

We demonstrate the working of our framework for two different RD systems (a Brusselator and a chemotactic model) with two different boundary conditions (zero Dirichlet and zero Neumann boundary conditions respectively). We validated our framework for these systems against known results in the literature, and experimented with the geometry of the underlying domains to obtain new results. In particular for Murray's chemotactic model, we perform experiments with developable and non-developable distortions of the reference rectangular domain. Our experiments show interesting variations in the emergent patterns due to non-developable distortions. We also apply Murray's model to study emergent patterns and branches for an arbitrary surface representing a gecko lizard. We upsample results to several higher levels with resolution upto one million FEM nodes, depending on the complexity of the pattern. We conclude that our framework can be used effectively to study emergent patterns and pattern branches for RD systems with or without cross diffusion, acting on arbitrary surfaces, with up to \dsdk{$\mathcal{O}(10^6)$}\FixRevThree{a few million} FEM nodes. Our framework is flexible, highly parallelisable and can be configured to use higher-order FEMs as well.

\bibliographystyle{spbasic}
\bibliography{bifurcation}

\let\noopsort\undefined
\let\printfirst\undefined
\let\singleletter\undefined
\let\switchargs\undefined
\appendix
\section*{Sample Photographs Sourced from Internet}
\emph{For Figure11}\vspace{0.2cm}\\
\noindent Images in (a) are from~\citesec{Gecko1},~\citesec{Gecko2},~\citesec{Gecko3},~\citesec{Gecko4} and \citesec{Gecko5}.

\bibliographystylesec{spbasic}
\bibliographysec{bifurcation}

\end{document}